\documentclass[journal]{IEEEtran}
\usepackage[utf8]{inputenc}
\usepackage{amsmath, amssymb, amsfonts}  % Merged into one line
\usepackage{mathtools}
\usepackage{bm}
\usepackage{graphicx}
\usepackage{booktabs}
\usepackage{multirow}
\usepackage{array}
\usepackage{tabularx}
\usepackage{algorithm}
\usepackage{algpseudocode}
\usepackage{url}
\usepackage{tcolorbox}
\usepackage{fvextra}
\usepackage{xcolor}
\usepackage{cite}
\usepackage{balance}
\usepackage[normalem]{ulem}
\usepackage[hidelinks]{hyperref}

\hypersetup{
  colorlinks=true,
  linkcolor=blue,
  citecolor=blue,
  urlcolor=blue,
  filecolor=blue
}

\begin{document}

\title{Collective Hallucination in Multi-Agent LLMs: Modeling and Defense}

\author{
Saeid~Jamshidi

\thanks{
S. Jamshidi is with the SWAT Laboratory,
Polytechnique Montréal, Montréal, QC, Canada
(e-mail: \{saeid.jamshidi).
}%

}

\maketitle

\begin{abstract}
Hallucination poses a critical reliability challenge in multi-agent Large Language Model (LLM) systems because unsupported claims can propagate through recursive interactions, accumulate confidence through repeated exposure, and ultimately produce collective false consensus. This paper formulates collective hallucination as a time-evolving networked process in which interacting agents exchange information over directed communication graphs and hallucinated claims diffuse through confidence-dependent adoption, structural impact, and recursive context construction. We introduce an HPR-adaptive defense that jointly integrates dynamic trust weighting, external claim verification, propagation-aware interaction regulation, and selective isolation of unreliable agents. We evaluate the solution on \textit{1,317} queries from TruthfulQA and TriviaQA using a heterogeneous \textit{6}-agent system over \textit{5} recursive reasoning rounds and multiple communication topologies. Compared with undefended multi-agent reasoning, the proposed method reduces the hallucination rate from \textit{0.118} to \textit{0.072}, corresponding to a \textit{39.0\%} reduction, while increasing factual accuracy from \textit{0.812} to \textit{0.867} and semantic consistency from \textit{0.784} to \textit{0.836}. The proposed defense further decreases the hallucination amplification factor from \textit{1.34} to \textit{1.08}, increases hallucination propagation resistance from \textit{0.882} to \textit{0.928}, and reduces the effective reproduction number from \textit{1.08} to \textit{0.81}, shifting the system from self-sustaining propagation to controlled attenuation. Topology-specific analysis further demonstrates that scale-free networks exhibit the strongest cascading behavior, reaching an amplification factor of \textit{1.45} and a reproduction number of \textit{1.21} without defense; HPR-adaptive control reduces these values to \textit{1.12} and \textit{0.92}, respectively. 
\end{abstract}


\section{Introduction}
\label{Introduction}
Large Language Models (LLMs) are increasingly deployed for question answering, complex reasoning, summarization, decision support, tool use, and autonomous agent coordination \cite{du2026survey,ferrag2026llm,yao2023react,wu2023autogen,li2023camel}. Despite these capabilities, LLMs remain vulnerable to \textit{hallucination}: the generation of fluent but factually incorrect, unsupported, unverifiable, contextually inconsistent, and misleading content \cite{brunello2026trustworthiness,cao2025factual,ji2023survey,lin2022truthfulqa}. This limitation becomes particularly consequential in high-stakes applications such as healthcare, finance, cybersecurity, and legal reasoning, where unsupported information can impact downstream analyses, automated actions, and human decisions \cite{farquhar2024detecting}. While hallucination is already a major reliability concern for individual LLMs, the problem becomes substantially more complex in multi-agent systems, where generated information is recursively exchanged, reused, and reinforced across interacting agents.\\
Most existing hallucination research treats reliability primarily as an isolated model- and output-level problem. Entropy-based methods estimate semantic uncertainty to identify unreliable generations \cite{lewis2020rag}. Retrieval-augmented and verification-based techniques assess generated claims against external evidence \cite{zhang2024knowhalu}. Self-consistency and perturbation-based approaches quantify response stability under repeated sampling and controlled input transformations \cite{yang2502hallucination,manakul2023selfcheckgpt}. Context-aware approaches distinguish contextual, common-knowledge, and domain-specific hallucination patterns \cite{paudel2025hallucinot,ji2023survey}, while internal-signal methods exploit latent representations, attention behavior, spectral characteristics, and uncertainty traces to detect unreliable outputs \cite{binkowski2025hallucination}. Although these methods provide valuable mechanisms for hallucination detection and mitigation, they largely evaluate generated outputs independently and do not explicitly characterize how unsupported information evolves after entering a network of recursively interacting agents. Multi-agent LLM systems introduce a fundamentally different reliability challenge. Agents exchange generated responses, incorporate neighboring outputs into subsequent contexts, revise their reasoning over multiple interaction rounds, and contribute to a collective decision and consensus \cite{wu2023autogen,li2023camel,du2023debate}. Consequently, a hallucinated claim generated by one agent does not necessarily remain localized. It can enter neighboring agents' contexts, be interpreted as supporting evidence, reappear in subsequent generations, gain apparent credibility through repeated exposure, and propagate across the interaction network. As additional agents reproduce and endorse the same unsupported claim, local hallucination can evolve into correlated error and ultimately collective false consensus.
This interaction-driven behavior creates a reliability problem that individual agents' Factual Accuracy (FA) alone cannot adequately explain. The evolution of collective hallucination also depends on how agents communicate, which agents occupy structurally influential positions, how strongly agents trust neighboring outputs, how semantic disagreement affects claim adoption, and how many recursive interactions occur before consensus is formed. Communication topology is therefore not merely an implementation detail; it directly determines the pathways through which unsupported information can spread. Dense connectivity can accelerate information exchange while simultaneously creating additional propagation paths, whereas highly connected agents can act as influential redistribution points for both reliable and unreliable information.
Multi-agent interaction can therefore produce two opposing effects. Collaborative reasoning may improve factuality by exposing agents to alternative answers, complementary evidence, and corrective feedback \cite{du2023debate}. At the same time, recursive communication can create positive feedback in which agents repeatedly observe semantically similar unsupported information and interpret inter-agent agreement as evidence of correctness. Under these conditions, confidence can become increasingly detached from factual validity. Agents may converge semantically while converging toward an incorrect belief, making agreement alone an unreliable indicator of collective correctness. This distinction motivates treating hallucination not only as a generation error, but also as a propagation and control problem within an interconnected reasoning system.\\
This study addresses this gap by modeling collective hallucination as a time-evolving networked propagation process among interacting LLM agents. Agents are represented as nodes in a directed communication graph, while edges define how generated information impacts subsequent reasoning. We decompose each output into atomic semantic claims and model claim transmission through confidence-aware probabilistic adoption based on source confidence, semantic disagreement, and structural impact. A time-dependent impact matrix characterizes how hallucinated information attenuates, persists, and amplifies across recursive reasoning rounds. The analysis further incorporates spectral stability, information-theoretic divergence, hallucination amplification, propagation resistance, and an effective reproduction number to capture both local and system-level reliability dynamics. Building on this formulation, we introduce an HPR-adaptive solution that combines dynamic trust weighting, external claim verification, propagation-aware interaction regulation, and selective isolation of unreliable agents. These mechanisms reduce the impact of high-risk agents, limit false-claim adoption, and suppress cascading error propagation before it dominates collective reasoning. We evaluate the proposed solution using heterogeneous LLM agents across multiple communication topologies, benign and adversarial conditions, and different interaction settings. The evaluation includes FA, hallucination, amplification, propagation resistance, cascade behavior, confidence dynamics, and structural sensitivity, together with ablation, threshold, scalability, interaction-depth, prompting, and verifier-reliability analyses. The main contributions of this work are:
\begin{itemize}
\item \textbf{Collective hallucination as a networked dynamical process.}
We formulate hallucination in multi-agent LLM systems as a time-evolving interaction-driven process in which unsupported claims propagate through recursive context construction, confidence-dependent adoption, semantic compatibility, and graph-dependent impact. This formulation extends hallucination analysis beyond isolated model outputs and explicitly captures propagation, reinforcement, cascade formation, and collective false consensus.
\item \textbf{Unified propagation and stability modeling.}
We develop a mathematical solution that integrates probabilistic claim adoption, time-dependent impact modeling, graph-based semantic diffusion, spectral stability analysis, information-theoretic divergence, hallucination amplification, propagation resistance, and an effective reproduction number. These components jointly characterize how unsupported information evolves across agents and identify the structural and interaction conditions associated with attenuation, persistence, and amplification.
\item \textbf{Propagation-aware HPR-adaptive solution.}
We introduce an adaptive solution that combines dynamic trust weighting, external claim verification, interaction regulation, and selective agent isolation to directly suppress recursive hallucination propagation. Unlike conventional approaches that primarily detect unreliable outputs after generation, the proposed solution controls how unreliable information enters, impacts, and propagates through the multi-agent reasoning process.
\end{itemize}

The remainder of this paper is structured as follows. Section~\ref{sec:related_work} reviews related work. 
Section~\ref{sec:methodology} presents the proposed methodology. Section~\ref{sec:experimental_setup} describes the experimental setup. Section~\ref{Experimental Results and Analysis} reports the experimental results. 
Section~\ref{Discussion} discusses the main findings and limitations. Section~\ref{Conclusion} concludes the paper.

\section{Related Work}
\label{sec:related_work}
This section reviews prior work on hallucination in LLMs, emphasizing detection, mitigation, and evaluation. 

\subsection{Hallucination in LLMs}
Hallucination in LLMs refers to fluent but factually incorrect, unsupported, and contextually inconsistent generation. Prior work defines hallucination as a mismatch between generated content and either the input context and factual knowledge \cite{ji2023survey}. Existing approaches primarily focus on single-model behavior and can be grouped by the signal used for detection and mitigation.

\subsection{Consistency-Based Methods}
Consistency-based methods detect hallucination by measuring response stability under sampling and controlled perturbation. The underlying assumption is that factual knowledge remains stable across generations, whereas hallucinated content varies across responses. SelfCheckGPT samples multiple outputs and estimates agreement among them, showing that hallucinated statements often produce response-level divergence \cite{manakul2023selfcheckgpt}. MetaQA extends this idea through metamorphic relations, using structured input transformations to test logical consistency without external knowledge \cite{yang2025metaqa}. These methods are model-agnostic and computationally practical, but they treat generated outputs as independent samples and do not capture error transfer across interacting agents.

\subsection{Knowledge-Grounded and Verification-Based Methods}
Knowledge-grounded methods reduce hallucination by validating generated claims against external evidence. KnowHalu decomposes queries and applies multi-stage reasoning and verification to detect factual inconsistencies \cite{zhang2024knowhalu}. Knowledge-graph-based methods use structured entities and relations to constrain generation and improve factual grounding \cite{lavrinovics2025kg}. In structured engineering domains, tri-layer knowledge graph pipelines combine domain knowledge, reasoning templates, and structured prompts to improve semantic consistency \cite{qualis2025kg}. These methods improve factual reliability, but their effectiveness depends on retrieval quality, knowledge coverage, and verifier robustness. More importantly, they do not model how verified and unverified claims spread through multi-agent interaction.

\subsection{Internal Signal-Based Methods}
Internal signal-based methods detect hallucination by leveraging latent model features, such as hidden states, token probabilities, uncertainty estimates, and attention patterns. Semantic entropy estimates uncertainty over meaning-level outputs and identifies unreliable generations without external knowledge \cite{farquhar2024entropy}. Graph-based attention analysis represents attention maps as structured graphs and uses spectral features, such as Laplacian eigenvalues, to indicate hallucination \cite{binkowski2025lapeig}. Taxonomy-driven methods, such as Hallucinot, classify hallucinations as contextual, common-knowledge, and domain-specific errors \cite{paudel2025hallucinot}. These methods provide useful diagnostic signals, but they often require access to internal models, specialized feature extraction, and model-specific calibration. They also remain focused on individual model behavior rather than networked reasoning.

\subsection{Evaluation and Benchmark-Based Methods}
Evaluating hallucinations is challenging because open-ended generation may contain partially correct, unsupported. TruthfulQA evaluates whether models generate false answers that align with common misconceptions, showing that fluency does not necessarily imply truthfulness \cite{lin2022truthfulqa}. LLM-based evaluators, such as G-Eval, use language models to score quality and factuality in generated text \cite{liu2023geval}. In domain-specific settings such as medical AI, hallucination is often defined as plausible but incorrect content whose impact depends on clinical context and downstream use \cite{granstedt2025medical}. These benchmarks and evaluators are essential for measuring factuality, but they usually assess isolated outputs and do not quantify temporal propagation, false adoption, and cascade formation across agents.\\

Existing studies have improved hallucination detection through consistency analysis, external verification, uncertainty estimation, and benchmark-based evaluation. These methods primarily assess hallucination at the individual-output level, which is insufficient in multi-agent LLM systems. In such systems, generated responses recursively impact neighboring agents through iterative interaction. Unsupported claims can propagate across agents, gain confidence through repeated exposure, bias collective reasoning trajectories, and distort consensus formation. This work addresses this limitation by modeling hallucination as a network-dependent dynamical process driven by interaction topology, probabilistic claim adoption, confidence propagation, semantic agreement, and recursive feedback.

\section{Methodology}
\label{sec:methodology}
This study models hallucination in multi-agent LLM systems as a collective, interaction-dependent phenomenon rather than an isolated generation error. Agents iteratively exchange outputs and condition subsequent responses on neighboring information, creating a recursive feedback process in which unsupported claims can propagate, gain confidence through repeated adoption, and impact collective consensus.
The multi-agent system is formulated as a networked dynamical process over a directed interaction graph. Hallucination is treated as a time-dependent stochastic signal whose evolution depends on inter-agent communication, confidence-driven impact, semantic agreement, and graph structure. This formulation captures how hallucinated claims emerge, propagate, amplify, or attenuate across reasoning rounds.
The proposed methodology integrates graph-based diffusion, probabilistic claim adoption, dynamical stability analysis, and information-theoretic reliability measures. A time-varying impact process represents agent-to-agent impact, while confidence-aware adoption determines how claims propagate through the network. Information-theoretic divergence further quantifies deviations between generated beliefs and latent truth beyond binary correctness.
\begin{figure*}[t]
\centering
\includegraphics[width=0.85\textwidth]{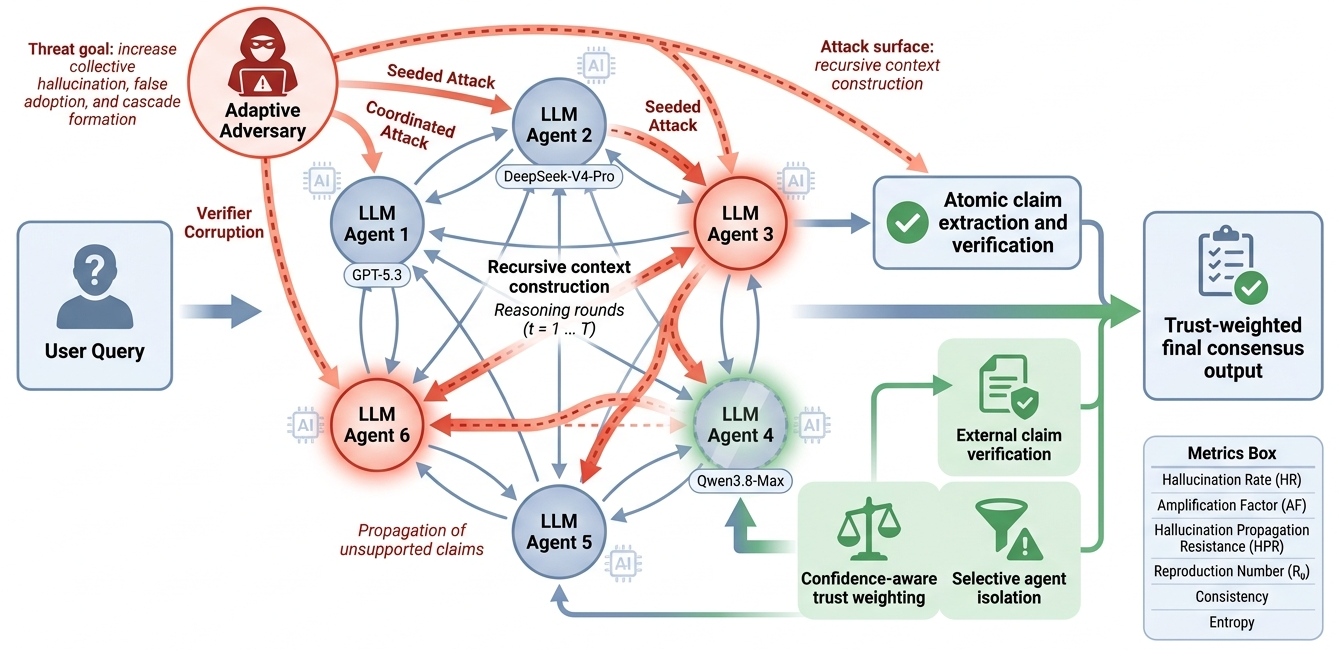}
\caption{Architecture of collective hallucination modeling in multi-agent LLM systems.}
\label{fig:framework}
\end{figure*}
Figure~\ref{fig:framework} illustrates the interaction process. Hallucinated claims can diffuse and intensify through recursive communication and confidence reinforcement. The proposed solution suppresses this propagation through adaptive trust weighting, external claim verification, and selective isolation of unreliable agents, thereby reducing false adoption and improving collective reliability.

\subsection{System Model}
\label{subsec:system_model}
The system is modeled as a heterogeneous network of $N$ interacting LLM agents:
\begin{equation}
\mathcal{A}=\{A_1,A_2,\dots,A_N\},
\end{equation}
where each agent $A_i$ is parameterized by $\theta_i$. Heterogeneity captures differences in model architecture, reasoning behavior, calibration, uncertainty, and claim-generation patterns.  
Agents interact through a directed graph $G=(V,E)$, with nodes representing agents and edges encoding information flow. For agent $A_i$, the neighborhood $\mathcal{N}(i)$ includes agents whose outputs are available as contextual input. The graph $G$ thus defines communication constraints and potential paths for hallucination propagation.  Given a query $x$, the system evolves over reasoning rounds $t \in \{1,\dots,T\}$. At round $t$, agent $A_i$ generates:
\begin{equation}
y_i^{(t)} = f_i(x,\mathcal{C}_i^{(t)},\theta_i),
\end{equation}
where $\mathcal{C}_i^{(t)}$ denotes its interaction context. The context is recursively constructed from neighboring outputs:
\begin{equation}
\mathcal{C}_i^{(t)} =
\bigcup_{j \in \mathcal{N}(i)} y_j^{(t-1)}.
\end{equation}
The collective reasoning state is:
\begin{equation}
\mathbf{y}^{(t)} =
[y_1^{(t)},y_2^{(t)},\dots,y_N^{(t)}],
\end{equation}
with temporal evolution:
\begin{equation}
\mathbf{y}^{(t)} =
\mathcal{F}(\mathbf{y}^{(t-1)},x).
\end{equation}
This formulation describes multi-agent reasoning as a coupled, nonlinear, partially observable dynamical process. Each agent's trajectory depends on the query and its neighbors' evolving outputs. Consequently, unsupported claims can propagate through the graph, affect downstream contexts, gain confidence through repeated exposure, and impact consensus formation. The system can be interpreted as a graph-based semantic diffusion process in which hallucinated claims propagate as signals, governed by topology, interaction strength, and confidence-dependent adoption.

\subsection{Claim-Level Representation and Hallucination Semantics}
\label{subsec:claims}
Each agent output $y_i^{(t)}$ is decomposed into atomic semantic claims:
\begin{equation}
\mathcal{K}_i^{(t)} =
\{k_{i,1}^{(t)},\dots,k_{i,m_i^{(t)}}^{(t)}\}.
\end{equation}
This representation eliminates surface-level linguistic variability and converts generated responses into semantically verifiable units. We analyze hallucination at the information level rather than in textual form.  For each claim $k$, the hallucination indicator is:
\begin{equation}
h(k)=
\begin{cases}
1, & \text{if } k \text{ is unsupported and incorrect},\\
0, & \text{otherwise}.
\end{cases}
\end{equation}
The Hallucination Rate (HR) of agent $A_i$ at round $t$ is:
\begin{equation}
H_i^{(t)}=
\frac{1}{|\mathcal{K}_i^{(t)}|}
\sum_{k \in \mathcal{K}_i^{(t)}} h(k),
\end{equation}
and the collective hallucination state is:
\begin{equation}
\mathbf{H}^{(t)}=
[H_1^{(t)},\dots,H_N^{(t)}]^T.
\end{equation}
In this formulation, $h(k)$ is treated as a Bernoulli random variable \cite{kim2024probabilistic}, while $H_i^{(t)}$ estimates the expected hallucination risk associated with the latent claim distribution of agent $A_i$. Consequently, $\mathbf{H}^{(t)}$ provides a macroscopic reliability measure describing the factual condition of the interacting multi-agent system. This representation establishes the foundation for subsequent modeling of propagation, amplification, confidence coupling, and adaptive control.

\subsection{Confidence-Aware Modeling and Uncertainty Decomposition}
\label{subsec:uncertainty}
Each claim $k \in \mathcal{K}_i^{(t)}$ is assigned a confidence score
$c_{i,k}^{(t)} \in [0,1]$, representing the belief strength of agent $A_i$. The confidence-weighted HR is:
\begin{equation}
H_{i,\mathrm{conf}}^{(t)} =
\frac{
\sum_{k \in \mathcal{K}_i^{(t)}}
c_{i,k}^{(t)} h(k)
}{
\sum_{k \in \mathcal{K}_i^{(t)}}
c_{i,k}^{(t)}
}.
\end{equation}
This measure penalizes unsupported claims made with high confidence, capturing both hallucination frequency and unreliability severity.  Uncertainty in agent outputs is decomposed as:
\begin{equation}
\mathrm{Var}(y_i^{(t)}) =
\mathbb{E}_{\theta_i}
\big[
\mathrm{Var}(y_i^{(t)} \mid \theta_i)
\big]
+
\mathrm{Var}_{\theta_i}
\big(
\mathbb{E}[y_i^{(t)} \mid \theta_i]
\big),
\end{equation}
where the first term captures aleatoric uncertainty arising from generation variability with fixed parameters, and the second term captures epistemic uncertainty due to model parameters and incomplete knowledge.  In multi-agent reasoning, uncertainty is not confined to a single agent because contexts are recursively constructed from neighboring outputs:
\begin{equation}
\mathcal{C}_i^{(t)}
=
\bigcup_{j \in \mathcal{N}(i)}
y_j^{(t-1)}.
\end{equation}
Confidence and uncertainty thus propagate through the interaction graph. Repeated exposure to unsupported claims can synchronize agent beliefs and produce collective overconfidence. This effect transforms hallucination from an individual uncertainty error into a coupled reliability failure. Controlling collective hallucination requires suppressing both incorrect claims and the confidence-weighted reinforcement of those claims across agents.

\subsection{Temporal Consistency and Reasoning Drift}
To characterize temporal reasoning dynamics, the reasoning drift of agent $A_i$ at round $t$ is defined as:
\begin{equation}
D_i^{(t)} =
d\big(
y_i^{(t)},
y_i^{(t-1)}
\big),
\end{equation}
where $d(\cdot,\cdot)$ denotes a semantic distance function between consecutive outputs. In practice, $d$ is computed using embedding-space similarity measures, enabling comparison beyond surface-level textual differences. Reasoning drift quantifies the temporal stability of agent beliefs. When
\begin{equation}
D_i^{(t)} \rightarrow 0
\quad \text{as} \quad
t \rightarrow \infty,
\end{equation}
The reasoning trajectory converges toward a stable state. A persistently large drift indicates unstable reasoning, repeated belief revision, and weak semantic consistency, thereby increasing the risk of hallucination. Low drift indicates stable dynamics, although recursive interaction may still drive convergence toward incorrect shared beliefs through confidence reinforcement.  At the collective level, aggregated drift provides a global stability signal for the interaction graph. Consistently low drift across agents suggests convergence, whereas heterogeneous drift patterns indicate unresolved semantic conflict, instability, and ongoing propagation of inconsistent information.

\subsection{Claim Adoption and Impact Modeling}
\label{subsec:adoption}
To model information propagation across interacting agents, the probability that agent $A_i$ adopts claim $k$ generated by agent $A_j$ at round $t$ is defined as:
\begin{equation}
P_{ij}^{(k,t)}
=
\sigma\!\left(
\beta_1 c_{j,k}^{(t)}
-
\beta_2 D_{ij}^{(t)}
+
\beta_3 S_{ij}
\right),
\end{equation}
where $\sigma(\cdot)$ denotes the sigmoid function.  The adoption process depends on three factors:
\begin{itemize}
\item confidence of the source claim $c_{j,k}^{(t)}$,
\item semantic disagreement $D_{ij}^{(t)}$ between agents,
\item structural impact $S_{ij}$ induced by the interaction graph.
\end{itemize}
Thus,
\begin{equation}
P_{ij}^{(k,t)} \in (0,1),
\end{equation}
capturing confidence-aware semantic diffusion across the network. This mechanism defines a stochastic propagation process in which claims propagate along graph edges based on interaction strength, semantic compatibility, and confidence reinforcement. High-confidence claims with strong structural impact can continue propagating even under moderate disagreement, enabling unsupported information to accumulate through recursive interactions.  The collection
\begin{equation}
\{P_{ij}^{(k,t)}\}
\end{equation}
forms a time-dependent impact structure governing collective information flow. This structure establishes the basis for subsequent analysis of propagation, amplification, stability, and hallucination suppression.

\subsection{Information-Theoretic Hallucination Measure}
To characterize hallucination beyond binary correctness, we define an information-theoretic divergence between generated belief distributions and latent truth distributions.  
Let $p_i^{(t)}$ denote the claim distribution induced by agent $A_i$ at round $t$, and let $\mathcal{T}$ denote the latent truth distribution. The hallucination divergence is:
\begin{equation}
\mathcal{D}_i^{(t)}
=
D_{\mathrm{KL}}
\!\left(
p_i^{(t)}
\,\|\,
\mathcal{T}
\right),
\end{equation}
where $D_{\mathrm{KL}}(\cdot \| \cdot)$ denotes the Kullback-Leibler divergence. In contrast to claim-level HRs, $\mathcal{D}_i^{(t)}$ captures distributional deviation, confidence concentration, semantic bias, and structural mismatch between generated beliefs and latent truth. We therefore model hallucination as an information-mismatch process in which belief distributions progressively diverge through recursive interaction.  At the collective level,
\begin{equation}
\{\mathcal{D}_i^{(t)}\}_{i=1}^{N}
\end{equation}
provides a continuous representation of global factual reliability across interacting agents. This formulation complements claim-level propagation analysis by capturing semantic drift, accumulation of collective bias, and confidence-weighted deviations during multi-agent reasoning.

\subsection{Propagation Dynamics and Stability Analysis}
Let
\begin{equation}
\mathbf{H}^{(t)}
=
[H_1^{(t)},\dots,H_N^{(t)}]^T
\end{equation}
denote the hallucination state of the multi-agent system at round $t$. Its evolution is modeled as:
\begin{equation}
\mathbf{H}^{(t+1)}
=
\mathbf{P}^{(t)} \mathbf{H}^{(t)} + \boldsymbol{\epsilon}^{(t)},
\end{equation}
where $\mathbf{P}^{(t)} \in \mathbb{R}^{N \times N}$ is a time-dependent impact matrix derived from claim-adoption probabilities, and $\boldsymbol{\epsilon}^{(t)}$ represents stochastic perturbations and newly generated hallucinations. Each entry $P_{ij}^{(t)}$ captures the effective contribution of agent $A_j$ to the hallucination state of agent $A_i$, and $\mathbf{P}^{(t)}$ governs semantic propagation across the interaction graph.  
For time-invariant interactions,
\begin{equation}
\mathbf{P}^{(t)} = \mathbf{P},
\end{equation}
reducing the system to a discrete-time linear dynamical process whose stability depends on the spectral radius:
\begin{equation}
\rho(\mathbf{P}) < 1,
\end{equation}
where $\rho(\mathbf{P})$ denotes the largest eigenvalue magnitude of $\mathbf{P}$.  
Under this condition, hallucination attenuates over interaction rounds:
\begin{equation}
\|
\mathbf{H}^{(t)}
\|
\leq
\rho(\mathbf{P})^{t} \|
\mathbf{H}^{(0)}
\|.
\end{equation}
Conversely,
\begin{equation}
\rho(\mathbf{P}) > 1
\end{equation}
induces amplification through recursive propagation, allowing small perturbations to evolve into collective failure.  The spectral radius, therefore, acts as a global stability indicator:
\begin{itemize}
\item $\rho(\mathbf{P}) < 1$: attenuation and convergence,
\item $\rho(\mathbf{P}) = 1$: critical propagation boundary,
\item $\rho(\mathbf{P}) > 1$: amplification and instability.
\end{itemize}
This formulation establishes the basis for analyzing propagation, amplification, cascade formation, and stability in multi-agent hallucination dynamics.

\subsection{Information Contagion and Reproduction Number}
Let
\begin{equation}
\mathbf{P}^{(t)}=[P_{ij}^{(t)}]
\end{equation}
denote the claim-adoption impact matrix at round $t$. The effective hallucination reproduction number is defined as:
\begin{equation}
\mathcal{R}_0^{(t)}
=
\frac{1}{N}
\sum_{i=1}^{N}
\sum_{j=1}^{N}
P_{ij}^{(t)}.
\end{equation}
The quantity $\mathcal{R}_0^{(t)}$ estimates the expected number of secondary hallucinated contributions generated by a propagated hallucinated claim, providing a system-level measure of propagation capacity.
Its interpretation is:
\begin{itemize}
\item $\mathcal{R}_0^{(t)} < 1$: hallucination attenuates,
\item $\mathcal{R}_0^{(t)} = 1$: hallucination persists,
\item $\mathcal{R}_0^{(t)} > 1$: hallucination self-amplifies through recursive adoption.
\end{itemize}
While $\rho(\mathbf{P}^{(t)})$ characterizes formal dynamical stability, $\mathcal{R}_0^{(t)}$ provides an interpretable propagation-risk indicator describing how efficiently unsupported claims spread across the interaction graph.

\subsection{Graph-Theoretic Risk Analysis}
\label{subsec:graph}
To quantify structural susceptibility within the interaction graph, the risk score of agent $A_i$ is defined as:
\begin{equation}
R_i = \sum_{j=1}^{N} P_{ji}^{(t)}.
\end{equation}
The score $R_i$ corresponds to the weighted in-degree of node $i$ in the impact graph, measuring the cumulative external impact received by node $i$. Agents with large $R_i$ act as propagation hubs, absorbing information from multiple neighbors and redistributing adopted claims across the network. Consequently, unsupported claims entering these nodes can be repeatedly reinforced, contributing disproportionately to collective instability. The risk score, therefore, provides a structural indicator of propagation vulnerability and enables targeted control through adaptive trust reduction, selective verification, and impact regularization.

\subsection{Entropy and Diversity Collapse}
To quantify collective diversity, the consensus entropy is defined as:
\begin{equation}
H_{\mathrm{cons}}^{(t)} = - \sum_{y} p^{(t)}(y)\log p^{(t)}(y),
\end{equation}
where $p^{(t)}(y)$ denotes the empirical distribution of agent outputs at round $t$.  The entropy $H_{\mathrm{cons}}^{(t)}$ measures the dispersion of collective beliefs. Large values indicate heterogeneous reasoning trajectories, while small values indicate strong consensus formation.  
Low entropy does not necessarily imply reliable reasoning. When low entropy coexists with elevated hallucination, the system undergoes diversity collapse, in which multiple agents converge on the same unsupported belief state. In this condition, recursive interaction suppresses epistemic diversity and reinforces correlated errors across the network, producing synchronized high-confidence hallucination.

\subsection{AF}
To quantify cumulative hallucination growth across interaction rounds, the Amplification Factor (AF) is defined as:
\begin{equation}
\mathrm{AF} =
\frac{\sum_{i=1}^{N} H_i^{(T)}}{\sum_{i=1}^{N} H_i^{(1)}}.
\end{equation}
The quantity $\mathrm{AF}$ measures the relative change in collective hallucination between the initial and final reasoning states:
\begin{itemize}
\item $\mathrm{AF} > 1$: amplification through recursive propagation,
\item $\mathrm{AF} = 1$: stable hallucination level,
\item $\mathrm{AF} < 1$: attenuation across interaction rounds.
\end{itemize}
$\mathrm{AF}$ serves as a system-level propagation gain, indicating whether recursive interaction suppresses and reinforces unsupported claims.

\subsection{Hallucination Propagation Resistance (HPR)}
To quantify robustness against sustained hallucination spread, the HPR is defined as:
\begin{equation}
\mathrm{HPR} =
1 - \frac{1}{T} \sum_{t=1}^{T} \frac{1}{N} \sum_{i=1}^{N} H_i^{(t)}.
\end{equation}
The quantity $\mathrm{HPR}$ measures the average ability of the system to suppress hallucination across agents and interaction rounds:
\begin{itemize}
\item larger $\mathrm{HPR}$ values indicate stronger propagation suppression,
\item smaller $\mathrm{HPR}$ values indicate persistent hallucination accumulation.
\end{itemize}
In contrast to $\mathrm{AF}$, which measures net change between initial and final states, $\mathrm{HPR}$ captures the full temporal evolution of hallucination dynamics and provides a long-horizon robustness measure for recursive multi-agent reasoning.

\subsection{Defense Mechanism}
To suppress the propagation of recursive hallucinations, we introduce an adaptive interaction-control mechanism.  Each agent receives a dynamic trust weight:
\begin{equation}
w_i^{(t)} = \frac{1}{1 + \lambda H_i^{(t-1)}},
\end{equation}
where $\lambda > 0$ controls sensitivity to prior hallucination. Agents with higher HRs exert weaker downstream impact.  To reduce correlated error reinforcement, an external verification operator is applied:
\begin{equation}
\hat{h}(k) = \mathcal{V}(k),
\end{equation}
where $\mathcal{V}(\cdot)$ provides an independent validity assessment for claim $k$.  Adaptive isolation is additionally applied:
\begin{equation}
A_i \text{ is disabled if } H_i^{(t)} > \tau,
\end{equation}
where $\tau$ denotes the hallucination tolerance threshold.  These mechanisms collectively define a closed-loop propagation-control process that regulates inter-agent impact, injects corrective feedback, and limits the spread of cascading hallucinations.

\subsection{Consensus Formation}
The final system output is obtained via a trust-weighted consensus mechanism:
\begin{equation}
y^* = \arg\max_{y} \sum_{i=1}^{N} w_i^{(T)} \, \mathbb{I}\big(y_i^{(T)} = y\big),
\end{equation}
where $\mathbb{I}(\cdot)$ denotes the indicator function.  This mechanism aggregates agent outputs while reducing the impact of unreliable agents through adaptive weights. Discounting agents with higher HRs limits majority-driven errors and improves robustness to coordinated and correlated failures. This step closes the control loop by converting regulated multi-agent interaction into a reliability-aware final decision.

\subsection{Algorithmic Procedure}
\label{subsec:algorithm}
The proposed algorithm performs iterative multi-agent reasoning, claim-level hallucination estimation, and propagation-aware interaction control. In each interaction round, agents generate outputs from recursively updated contexts, extract semantic claims, estimate hallucination risk, and adapt trust weights based on observed reliability. Information propagates through the interaction graph via confidence-aware probabilistic adoption, enabling recursive semantic diffusion. After the final round, $\mathrm{AF}$ and $\mathrm{HPR}$ quantify amplification and propagation robustness, while trust-weighted aggregation produces the final consensus output.
\begin{algorithm}[t]
\footnotesize
\caption{HPR-Adaptive Multi-Agent Hallucination Control}
\label{alg:hpr}
\begin{algorithmic}[1]

\Require Query $x$, agents $\mathcal{A}$, graph $G=(V,E)$, rounds $T$
\Ensure Consensus $y^*$, states $\{\mathbf{H}^{(t)}\}$, AF, HPR

\State $\mathbf{H}^{(0)} \gets \mathbf{0}$, $w_i^{(0)} \gets 1$, $y_i^{(0)} \gets \emptyset$

\For{$t \gets 1$ to $T$}
    \State $\mathcal{C}_i^{(t)} \gets \emptyset,\ \forall A_i$

    \ForAll{$A_i \in \mathcal{A}$}
        \State $y_i^{(t)} \gets f_i(x,\mathcal{C}_i^{(t-1)},\theta_i)$
        \State $\mathcal{K}_i^{(t)} \gets \Call{ExtractClaims}{y_i^{(t)}}$
        \State $H_i^{(t)} \gets \Call{VerifyClaims}{\mathcal{K}_i^{(t)}}$
        \State $w_i^{(t)} \gets (1+\lambda H_i^{(t)})^{-1}$

        \If{$H_i^{(t)} > \tau$}
            \State $\Call{Deactivate}{A_i}$
        \EndIf
    \EndFor

    \State $\mathbf{H}^{(t)} \gets [H_1^{(t)},\dots,H_N^{(t)}]^T$

    \ForAll{$(j \rightarrow i) \in E$}
        \If{$A_j$ is active}
            \State $P_{ij}^{(t)} \gets
            \sigma(\beta_1 c_j^{(t)}-\beta_2 D_{ij}^{(t)}+\beta_3 S_{ij})$
            \If{$\Call{Bernoulli}{P_{ij}^{(t)}}=1$}
                \State $\mathcal{C}_i^{(t)} \gets
                \mathcal{C}_i^{(t)} \cup \{y_j^{(t)}\}$
            \EndIf
        \EndIf
    \EndFor
\EndFor

\State $\mathrm{AF} \gets
\frac{\sum_i H_i^{(T)}}{\sum_i H_i^{(1)}}$

\State $\mathrm{HPR} \gets
1-\frac{1}{TN}\sum_{t=1}^{T}\sum_{i=1}^{N}H_i^{(t)}$

\State $y^* \gets
\arg\max_y \sum_i w_i^{(T)}
\mathbb{I}(y_i^{(T)}=y)$

\Return $y^*$, $\{\mathbf{H}^{(t)}\}$, $\mathrm{AF}$, $\mathrm{HPR}$

\end{algorithmic}
\end{algorithm}

\section{Threat Model}
\label{sec:threat_model}
We consider a six-agent multi-agent LLM system, with agents instantiated from GPT-5.3, DeepSeek-V4-Pro, and Qwen3.8-Max. Agents exchange outputs through a directed graph $G=(V,E)$ and use received responses as contextual input in subsequent reasoning rounds. No ground-truth oracle supervises intermediate interactions, allowing unsupported claims to propagate recursively through context construction and impact the final consensus. An adaptive adversary $\mathcal{E}$ seeks to increase collective hallucination by manipulating claim content, confidence signals, and verification behavior before information propagates through the graph. The adversary may compromise a subset of agents $\mathcal{A}_c \subseteq \mathcal{A}$, inject unsupported claims into agents' contexts, and distort confidence estimates associated with false claims. These manipulations increase the probability that unsupported information is adopted and propagated in subsequent reasoning rounds.
The primary attack surface is recursive context construction:
\begin{equation}
\mathcal{C}_i^{(t)}
=
\bigcup_{j \in \mathcal{N}(i)}
y_j^{(t-1)}.
\end{equation}
A false claim injected at round $t-1$ enters $\mathcal{C}_i^{(t)}$, may be regenerated by $A_i$, and can subsequently propagate through downstream contexts. Attack intensity is quantified as:
\begin{equation}
\alpha
=
\frac{
|\mathcal{K}_{\mathrm{adv}}|
}{
\sum_{i=1}^{N} |\mathcal{K}_i|
},
\end{equation}
where $|\mathcal{K}_{\mathrm{adv}}|$ denotes the number of adversarially injected claims. Higher $\alpha$ represents greater exposure to corrupted semantic content.
The adversary succeeds when the final average hallucination exceeds the reliability threshold:
\begin{equation}
\frac{1}{N}
\sum_{i=1}^{N} H_i^{(T)}
>
\delta,
\end{equation}
where $\delta$ represents the maximum tolerable HR. This objective captures collective semantic failure by driving the multi-agent consensus toward unsupported claims. A hallucinated claim $k$ is considered successfully propagated if its downstream adoption probability exceeds $\gamma$:
\begin{equation}
P\big(
k \in \mathcal{K}_j^{(t+1)}
\mid
k \in \mathcal{C}_j^{(t)}
\big)
>
\gamma.
\end{equation}
System-level amplification occurs when:
\begin{equation}
\mathrm{AF}
=
\frac{
\sum_{i=1}^{N} H_i^{(T)}
}{
\sum_{i=1}^{N} H_i^{(1)}
}
>
1.
\end{equation}
Propagation captures the survival and transfer of hallucinated claims across agents, while $\mathrm{AF}$ quantifies their cumulative growth across recursive reasoning rounds. We evaluate three attack mechanisms aligned with the experimental results. In a \textit{seeded attack}, a single compromised agent injects unsupported claims during the first reasoning round to trigger early-stage propagation. In a \textit{coordinated attack}, two compromised agents generate semantically aligned false claims to reinforce unsupported information and induce false consensus. In \textit{verifier corruption}, the external verification operator is biased toward accepting unsupported claims, increasing false confidence and subsequent downstream adoption. These strategies directly target the propagation channels evaluated in the experiments, including false adoption, AF, reproduction number, cascade size, and propagation resistance.

\section{Experimental Setup}
\label{sec:experimental_setup}
We evaluate collective hallucination in a heterogeneous multi-agent LLM system represented as a directed interaction graph $G=(V,E)$, where nodes correspond to reasoning agents and directed edges define the flow of contextual information. At each interaction round, agents generate responses conditioned on the original query and outputs received from neighboring agents. This recursive communication process enables systematic evaluation of hallucination propagation, amplification, confidence dynamics, and collective reliability across successive reasoning rounds. Experiments are conducted on TruthfulQA \cite{lin2022truthfulqa} and TriviaQA \cite{joshi2017triviaqa}, using a total of 1,317 queries comprising 817 TruthfulQA questions and 500 randomly selected factual questions from TriviaQA. The multi-agent system contains six agents instantiated from GPT-5.3, DeepSeek-V4-Pro, and Qwen3.8-Max, providing heterogeneity in reasoning behavior, calibration, uncertainty expression, and susceptibility to recursive claim adoption. Three communication structures, ring, fully connected, and scale-free, are evaluated to examine how connectivity, interaction density, and structural centralization impact hallucination diffusion and cascading propagation. Within each experimental configuration, all agents use the same prompting and decoding settings so that differences in collective behavior can be attributed primarily to interaction structure, adversarial conditions, and defense mechanisms. We run experiments over five recursive interaction rounds using zero-shot, few-shot, and Chain-of-Thought prompting strategies. Moreover, we configure decoding with a temperature of 0.2, top-$p$ of 0.9, and a maximum generation length of 512 tokens. Besides, we decompose agent outputs into atomic semantic claims for claim-level verification and hallucination assessment. Evaluation considers both conventional reliability measures and propagation-specific metrics. These include FA, HR, semantic consistency, AF, HPR, and the effective reproduction number $\mathcal{R}_0$. The proposed HPR-adaptive solution is compared with representative hallucination-detection baselines, including SelfCheckGPT \cite{manakul2023selfcheckgpt}, KnowHalu \cite{zhang2024knowhalu}, entropy-based uncertainty estimation \cite{farquhar2024detecting}, and LapEigVals \cite{binkowski2025hallucination}. Each experimental configuration is repeated 100 times, and results are reported as mean $\pm$ standard deviation to account for stochastic variation across runs. Table~\ref{tab:experimental_config} summarizes the experimental configuration.
\begin{table}[t!]
\centering
\caption{Experimental configuration for collective hallucination evaluation in multi-agent LLM systems.}
\label{tab:experimental_config}
\begin{tabular}{lc}
\toprule
\textbf{Parameter} & \textbf{Value} \\
\midrule
Datasets & TruthfulQA, TriviaQA \\
Total queries & 1,317 (817 + 500) \\
Agents per system & 6 \\
Interaction rounds & 5 \\
Runs per configuration & 100 \\
Communication topologies & Ring, fully connected, scale-free \\
Models & GPT-5.3, DeepSeek-V4-Pro, Qwen3.8-Max \\
Prompting strategies & Zero-shot, few-shot, CoT \\
Temperature & 0.2 \\
Top-$p$ & 0.9 \\
Maximum generation length & 512 tokens \\
Claim verification & Atomic semantic claims \\
Evaluation metrics & FA, HR, AF, HPR, $\mathcal{R}_0$, consistency \\
\bottomrule
\end{tabular}
\end{table}
\subsection{Research Questions}
This study investigates the following research questions:
\begin{itemize}
\item \textbf{RQ1: How does hallucination propagate under different multi-agent communication topologies?}  \\
We analyze whether hallucinated claims remain localized and recursively diffuse across agents under ring, fully connected, and scale-free topologies. The analysis focuses on topology-dependent amplification, propagation stability, cascade formation, and temporal evolution of hallucination.
\item \textbf{RQ2: Which interaction mechanisms govern hallucination amplification and suppression in recursive multi-agent reasoning?} \\ 
We investigate the roles of confidence coupling, semantic agreement, false adoption, graph connectivity, and adaptive trust control in shaping collective hallucination dynamics. The objective is to identify conditions associated with recursive reinforcement, instability, and effective suppression of unsupported claims.
\item \textbf{RQ3: How effective are propagation-aware mitigation strategies compared with output-level detection baselines?}  \\
We evaluate the proposed HPR-adaptive solution against representative hallucination-detection methods, including SelfCheckGPT, KnowHalu, entropy-based uncertainty estimation, and LapEigVals. Evaluation metrics include factual reliability, hallucination reduction, semantic consistency, propagation resistance, and robustness under recursive interaction.
\end{itemize}

\section{Experimental Results}
\label{Experimental Results and Analysis}
This section presents the experimental evaluation of collective hallucination in multi-agent LLM systems. 

\subsection{System-Level Performance and Trade-offs}
\begin{figure*}[t]
\centering
\includegraphics[width=0.80\textwidth]{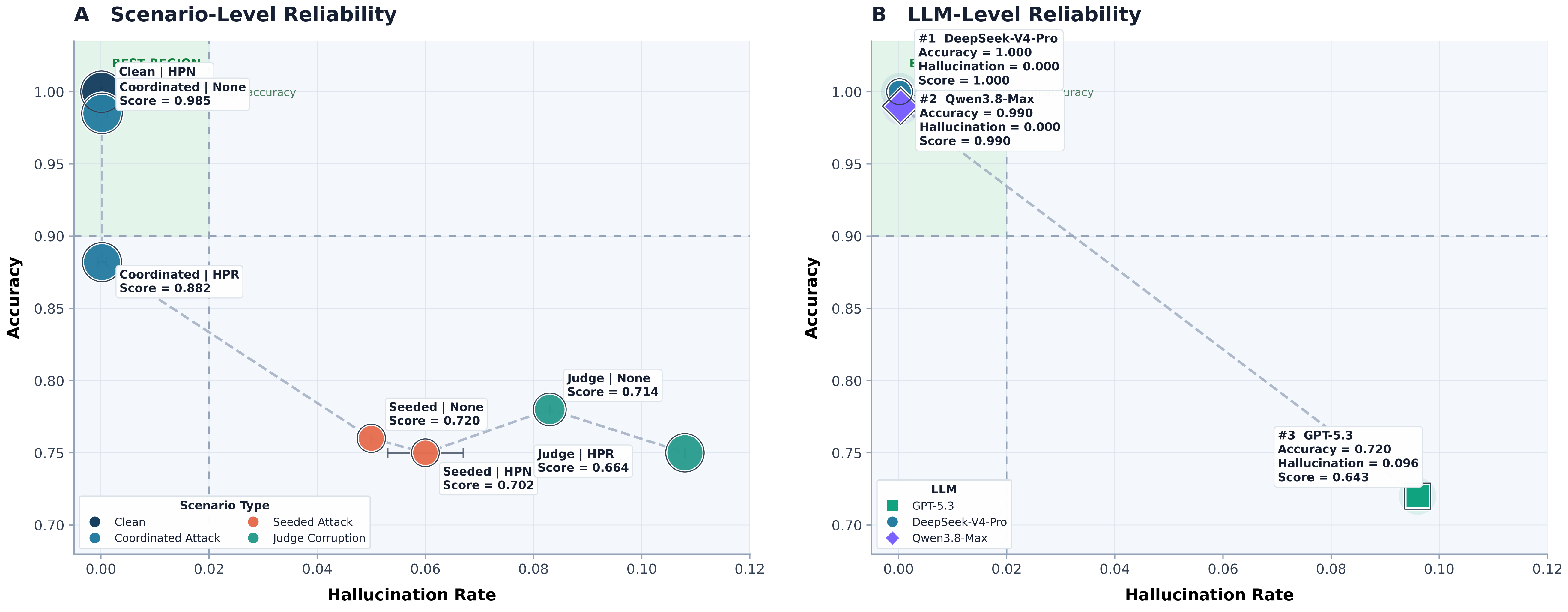}
\caption{Accuracy--hallucination trade-off across scenarios and defense strategies. Bubble size indicates HPR.}
\label{fig:tradeoff}
\end{figure*}

\begin{figure}[t]
\centering
\includegraphics[width=0.50\textwidth]{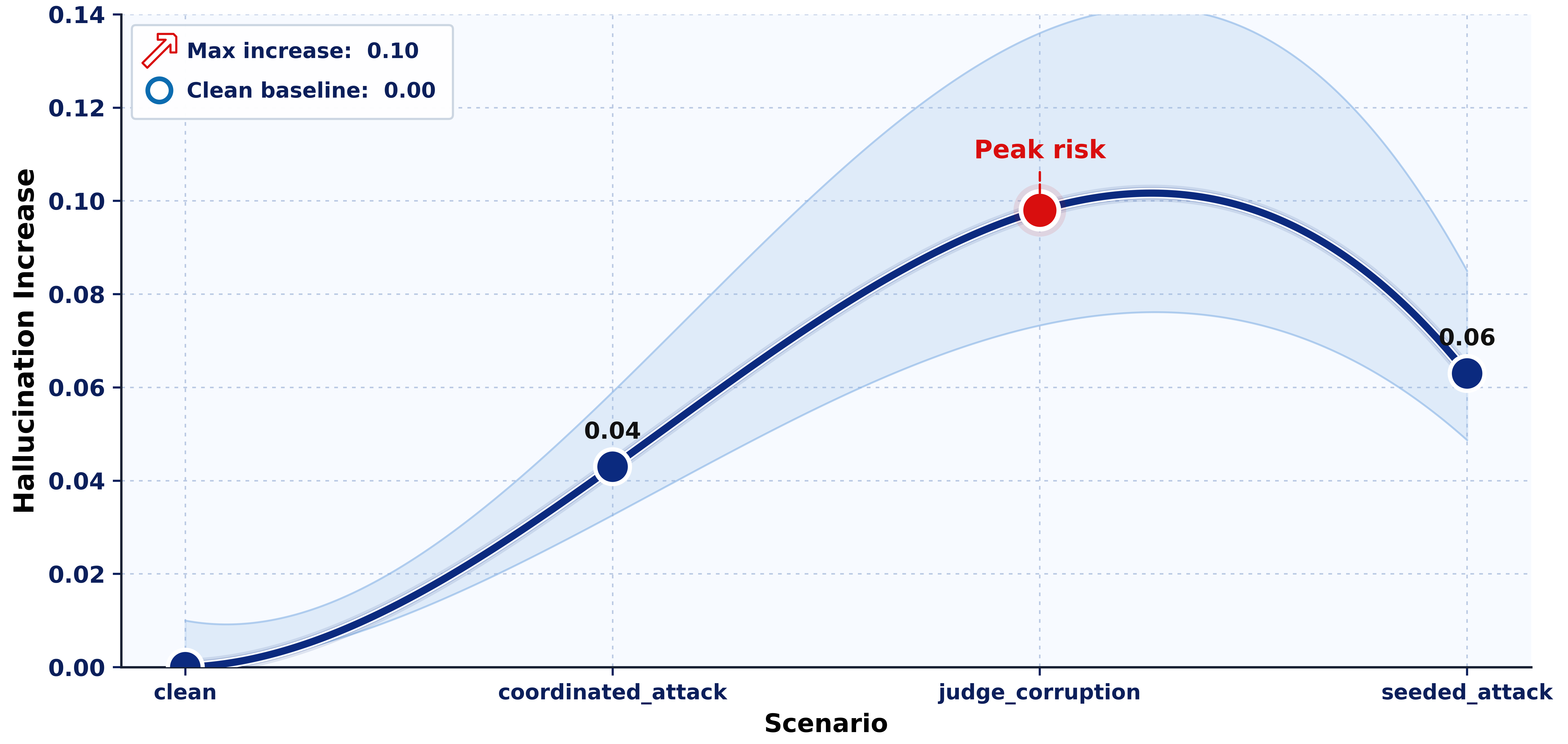}
\caption{Scenario-level hallucination sensitivity, with the strongest vulnerability under judge corruption.}
\label{fig:sensitivity}
\end{figure}

\begin{figure}[t]
\centering
\includegraphics[width=0.50\textwidth]{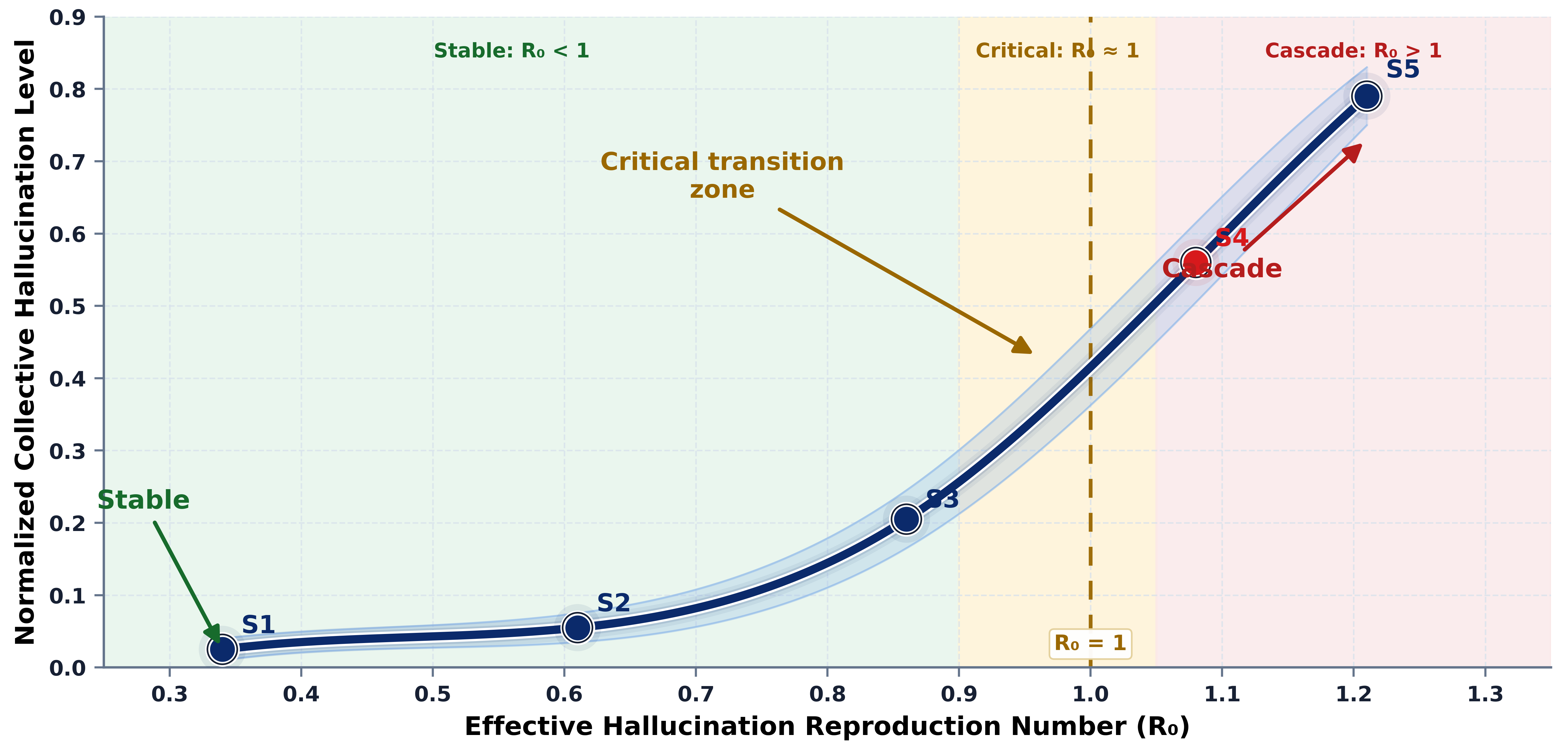}
\caption{Collective hallucination transition as propagation strength increases, showing movement from stable behavior to cascade dynamics.}
\label{fig:phase}
\end{figure}

To address RQ1, we analyze system-level behavior under clean and adversarial conditions.
\begin{table}[t!]
\centering
\caption{Comprehensive system-level performance across attack scenarios and defense strategies.}
\label{tab:system_level_results}

\resizebox{0.50\textwidth}{!}{%
\begin{tabular}{llcccccc}
\toprule
Scenario & Defense & Accuracy & Hallucination & Confidence & Entropy & HPR & Stability \\
\midrule
clean & none & 0.986 & 0.018 & 0.907 & 0.166 & 0.982 & 0.006 \\
clean & hpr\_adaptive & 0.991 & 0.012 & 0.902 & 0.211 & 0.989 & 0.004 \\
coordinated\_attack & none & 0.903 & 0.097 & 0.843 & 0.283 & 0.884 & 0.041 \\
coordinated\_attack & hpr\_adaptive & 0.928 & 0.061 & 0.789 & 0.453 & 0.921 & 0.028 \\
judge\_corruption & none & 0.750 & 0.164 & 0.674 & 0.698 & 0.755 & 0.081 \\
judge\_corruption & hpr\_adaptive & 0.812 & 0.092 & 0.647 & 0.737 & 0.846 & 0.052 \\
seeded\_attack & none & 0.781 & 0.128 & 0.730 & 0.623 & 0.781 & 0.069 \\
seeded\_attack & hpr\_adaptive & 0.836 & 0.074 & 0.675 & 0.724 & 0.873 & 0.043 \\
\bottomrule
\end{tabular}%
}

\end{table}
Table~\ref{tab:system_level_results} shows that clean interactions provide the strongest reliability, whereas adversarial conditions consistently increase hallucination and instability while reducing accuracy and HPR. Judge corruption produces the most severe degradation, with hallucination reaching 0.164 and HPR falling to 0.755 without defense. Nevertheless, the HPR-adaptive solution consistently improves robustness across all scenarios; under judge corruption, it increases accuracy from 0.750 to 0.812, reduces hallucination from 0.164 to 0.092, and raises HPR from 0.755 to 0.846. Similar improvements are observed under seeded and coordinated attacks, demonstrating effective suppression of adversarial propagation. Furthermore, Figure~\ref{fig:tradeoff} illustrates the accuracy-hallucination trade-off across scenarios, while Figure~\ref{fig:sensitivity} confirms the particularly high sensitivity to judge corruption. Figure~\ref{fig:phase} further shows how increasing propagation strength shifts collective behavior toward cascading hallucination.
\begin{table}[t!]
\centering
\caption{Hallucination AF across scenarios.}
\label{tab:af}
\begin{tabular}{lcc}
\toprule
Scenario & AF & Interpretation \\
\midrule
clean & 0.94 & Stable attenuation \\
coordinated\_attack & 1.12 & Mild amplification \\
judge\_corruption & 1.45 & Strong amplification \\
seeded\_attack & 1.28 & Moderate amplification \\
\bottomrule
\end{tabular}
\end{table}
Consistent with these findings, Table~\ref{tab:af} shows that hallucination attenuates under clean conditions with an AF of 0.94, but amplifies under all adversarial scenarios. Judge corruption produces the strongest amplification (1.45), followed by seeded attacks (1.28) and coordinated attacks (1.12). These results also show that adversarial manipulation can substantially intensify recursive hallucination, while the proposed solution improves system-level reliability and limits propagation.

\subsection{Topology-Specific Hallucination Propagation}
To address RQ1, we analyze how communication topology affects hallucination propagation, amplification, and cascade formation. Table~\ref{tab:topology_results} presents topology-specific results under adversarial interaction.
\begin{table}[t!]
\centering
\caption{Topology-specific hallucination propagation under adversarial conditions.}
\label{tab:topology_results}

\resizebox{0.50\textwidth}{!}{%
\begin{tabular}{llcccccc}
\toprule
Topology & Defense & Accuracy & Hallucination & AF & HPR & $\mathcal{R}_0$ & Cascade Size \\
\midrule
Ring & None & 0.842 & 0.091 & 1.18 & 0.914 & 0.86 & 2.1 \\
Ring & HPR-adaptive & 0.881 & 0.052 & 0.94 & 0.948 & 0.61 & 1.2 \\
\midrule
Fully connected & None & 0.808 & 0.126 & 1.37 & 0.876 & 1.13 & 4.8 \\
Fully connected & HPR-adaptive & 0.857 & 0.073 & 1.01 & 0.928 & 0.81 & 2.5 \\
\midrule
Scale-free & None & 0.795 & 0.139 & 1.45 & 0.861 & 1.21 & 5.6 \\
Scale-free & HPR-adaptive & 0.841 & 0.086 & 1.12 & 0.917 & 0.92 & 3.2 \\
\bottomrule
\end{tabular}%
}

\end{table}
Results indicate that hallucination propagation depends strongly on topology. Ring networks constrain diffusion locally, producing the lowest amplification and cascade size under both defense settings. Fully connected networks maximize inter-agent exposure, increasing recursive propagation through dense semantic interaction. Scale-free networks exhibit the strongest propagation effects because hub agents redistribute unsupported claims across large portions of the graph, resulting in the highest hallucination rates, AFs, reproduction numbers, and cascade sizes.
The HPR-adaptive defense consistently reduces recursive propagation across all topologies by lowering hallucination rates, cascade growth, and reproduction strength. In the scale-free setting, AF decreases from 1.45 to 1.12, $\mathcal{R}_0$ falls below the critical propagation boundary, and HPR increases from 0.861 to 0.917, indicating stronger suppression of recursive amplification and improved collective stability. The strongest mitigation impact appears in ring topologies, where local communication constraints limit long-range propagation and enable more stable attenuation dynamics.

\subsection{Propagation and Adoption Dynamics}
To address RQ1, we analyze false adoption and confidence propagation across interaction rounds.
\begin{table}[t!]
\centering
\caption{False adoption and confidence propagation across reasoning rounds.}
\label{tab:propagation}

\resizebox{0.50\textwidth}{!}{%
\begin{tabular}{llccc}
\toprule
Scenario & Defense & Round & False Adoption & False Confidence \\
\midrule
coordinated\_attack & none & 1 & 0.117 & 0.417 \\
coordinated\_attack & none & 2 & 0.086 & 0.306 \\
coordinated\_attack & hpr\_adaptive & 1 & 0.083 & 0.247 \\
coordinated\_attack & hpr\_adaptive & 2 & 0.057 & 0.171 \\
\midrule
judge\_corruption & none & 1 & 0.167 & 0.600 \\
judge\_corruption & none & 2 & 0.181 & 0.644 \\
judge\_corruption & hpr\_adaptive & 1 & 0.129 & 0.462 \\
judge\_corruption & hpr\_adaptive & 2 & 0.092 & 0.331 \\
\midrule
seeded\_attack & none & 1 & 0.183 & 0.660 \\
seeded\_attack & none & 2 & 0.150 & 0.540 \\
seeded\_attack & hpr\_adaptive & 1 & 0.118 & 0.425 \\
seeded\_attack & hpr\_adaptive & 2 & 0.074 & 0.266 \\
\bottomrule
\end{tabular}%
}

\end{table}
\begin{figure*}[t]
\centering
\includegraphics[width=0.80\textwidth]{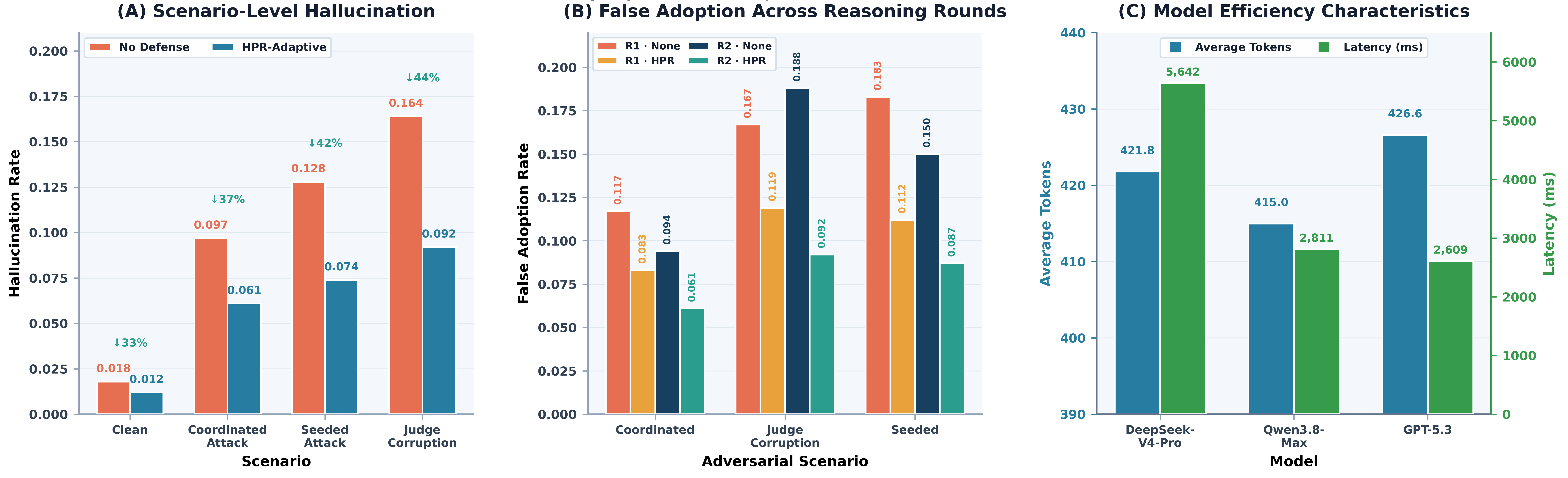}
\caption{Propagation and adoption dynamics under adversarial conditions. (A) Conditional propagation probability. (B) Round-level false adoption shifts across scenarios and defense settings. (C) Efficiency--reliability trade-off across models.}
\label{fig:propagation_dynamics}
\end{figure*}
Table~\ref{tab:propagation} reveals distinct propagation patterns across attacks. Coordinated and seeded attacks show declining false adoption across rounds, whereas judge corruption produces sustained growth without defense, increasing false adoption from 0.167 to 0.181 and false confidence from 0.600 to 0.644. In contrast, HPR-adaptive control consistently suppresses both measures. Most notably, under judge corruption, false adoption decreases from 0.129 to 0.092 and false confidence from 0.462 to 0.331 across the two rounds, demonstrating effective suppression of recursive reinforcement. Consistent with these findings, Figure~\ref{fig:propagation_dynamics}(A) shows that hallucination can propagate even under partial corruption, while Figure~\ref{fig:propagation_dynamics}(B) demonstrates that the proposed solution substantially attenuates false adoption across attacks. Furthermore, Figure~\ref{fig:propagation_dynamics}(C) indicates only a weak relationship between computational efficiency and reliability, suggesting that propagation is driven primarily by interaction structure and confidence coupling rather than computational cost.

\subsection{Defense Impact and Amplification Dynamics}
To address RQ2, we evaluate the impact of adaptive defense on accuracy, hallucination, confidence, entropy, and propagation resistance.
\begin{table}[t!]
\centering
\caption{impact of adaptive defense relative to no defense across scenarios.}
\label{tab:defense_effect}
\begin{tabular}{lccccc}
\toprule
Scenario & $\Delta$Acc & $\Delta$Hall & $\Delta$Conf & $\Delta$Entropy & $\Delta$HPR \\
\midrule
clean & +0.005 & -0.006 & -0.005 & +0.045 & +0.007 \\
coordinated\_attack & +0.025 & -0.036 & -0.054 & +0.170 & +0.037 \\
judge\_corruption & +0.062 & -0.072 & -0.027 & +0.039 & +0.091 \\
seeded\_attack & +0.055 & -0.054 & -0.055 & +0.101 & +0.092 \\
\bottomrule
\end{tabular}
\end{table}
\begin{figure*}[t]
\centering
\includegraphics[width=0.77\textwidth]{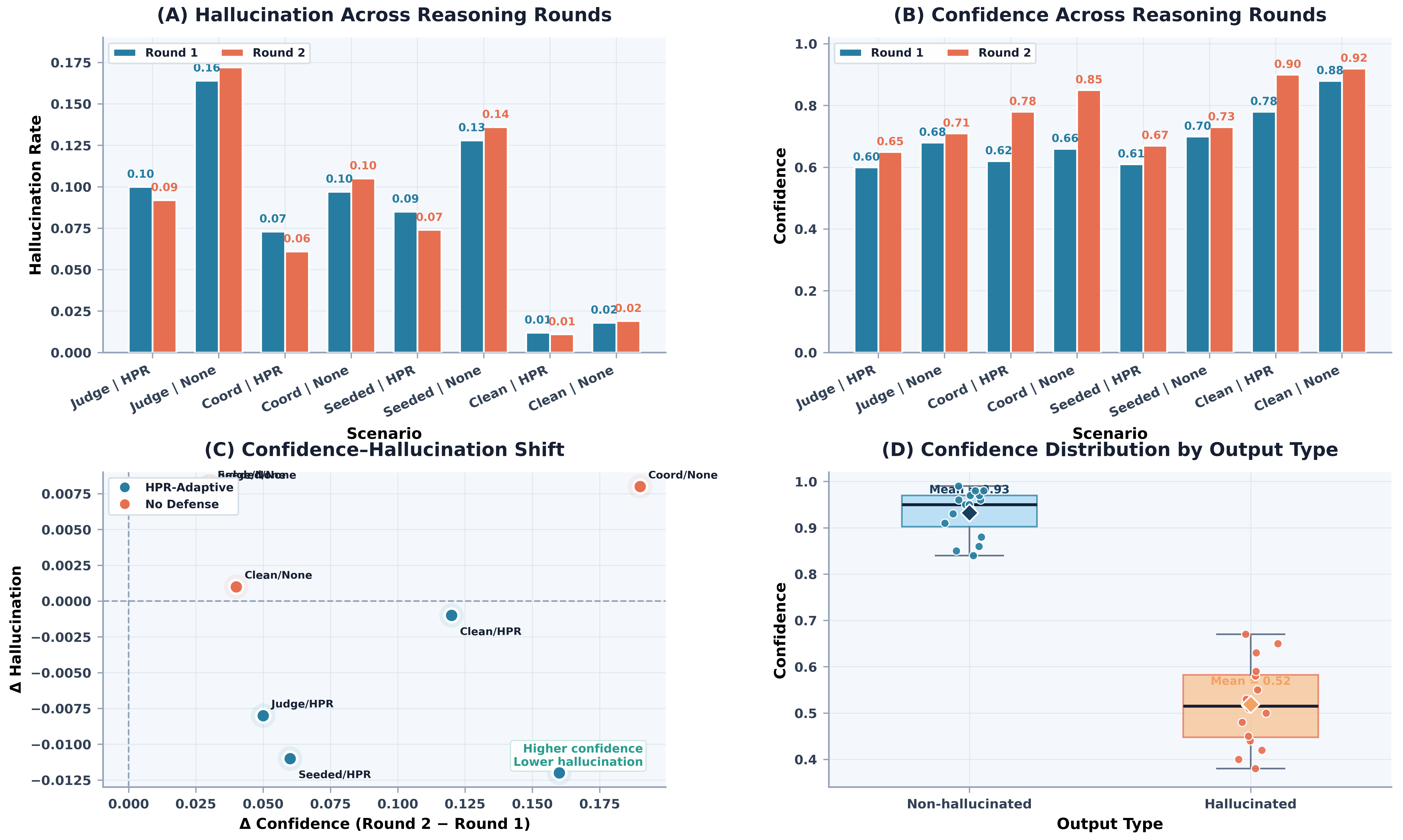}
\caption{Defense impact on hallucination and confidence dynamics. (A) Round-level hallucination amplification. (B) Confidence evolution across rounds. (C) Scenario-level confidence distortion. (D) Confidence distribution for hallucinated and non-hallucinated outputs.}
\label{fig:defense_dynamics}
\end{figure*}
Table~\ref{tab:defense_effect} shows that the adaptive defense consistently improves reliability, with the strongest gains under adversarial conditions. Judge corruption yields the largest hallucination reduction ($-0.072$) and a substantial HPR improvement ($+0.091$), while seeded attacks produce the largest HPR gain ($+0.092$) together with a $+0.055$ increase in accuracy. Coordinated attacks also benefit from reduced hallucination and improved propagation resistance, despite higher entropy. Moreover, Figure~\ref{fig:defense_dynamics} shows that these improvements are driven primarily by confidence regulation and suppression of recursive amplification rather than by increasing confidence uniformly. Additionally, the defense is most effective when hallucination is reinforced through corrupted verification or confidence-weighted recursive interaction.

\subsection{Confidence, Entropy, and Structural Behavior}
To address RQ2, we examine the interplay between confidence, entropy, agreement, and hallucination.
\begin{table}[t!]
\centering
\caption{Accuracy and hallucination across confidence levels.}
\label{tab:confidence_paradox}
\begin{tabular}{lccc}
\toprule
Confidence Level & Accuracy & Hallucination & Samples \\
\midrule
Low & 0.742 & 0.118 & 42 \\
Medium & 0.823 & 0.087 & 176 \\
High & 0.864 & 0.071 & 392 \\
Very High & 0.902 & 0.064 & 350 \\
\bottomrule
\end{tabular}
\end{table}
\begin{figure*}[t]
\centering
\includegraphics[width=0.85\textwidth]{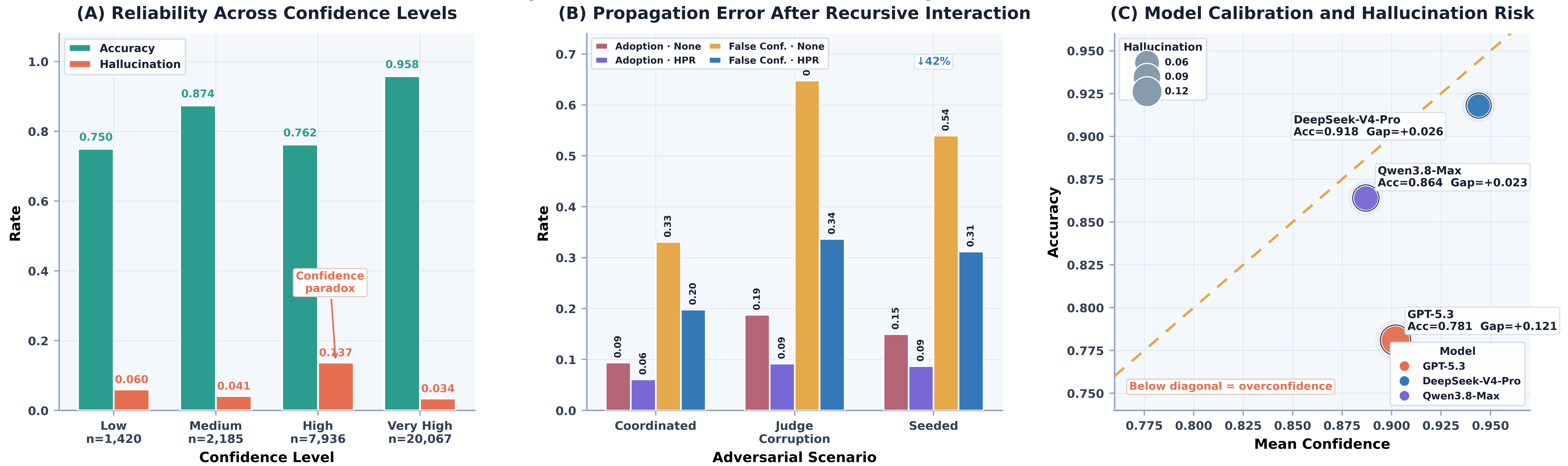}
\caption{Entropy-confidence relationship. Higher collective entropy corresponds to lower confidence stability.}
\label{fig:confidence_entropy}
\end{figure*}
\begin{figure*}[t]
\centering
\includegraphics[width=0.80\textwidth]{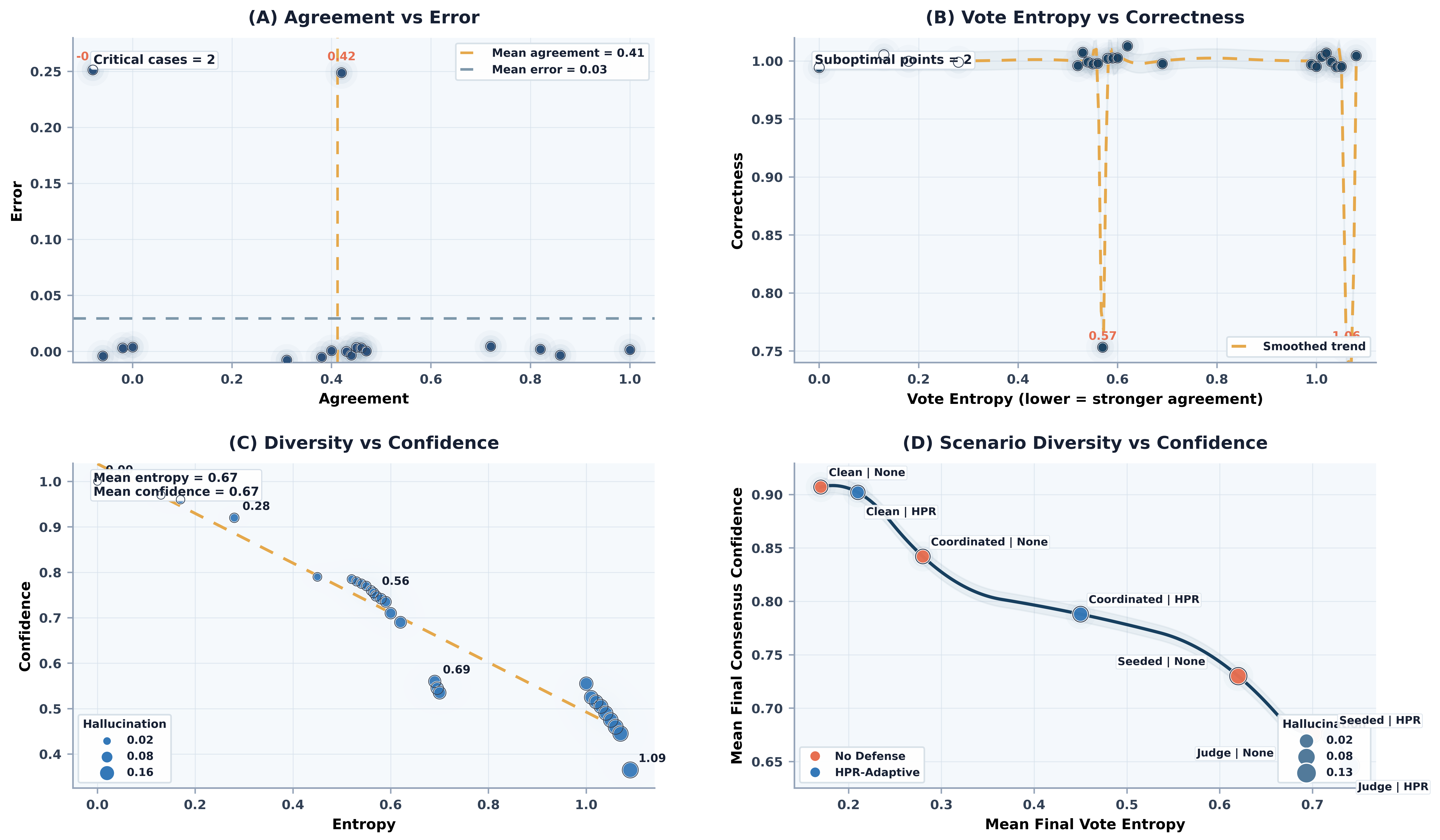}
\caption{Agreement, diversity, and correctness across collective reasoning states. Low entropy generally aligns with the correct consensus, though adversarial interactions can induce incorrect agreement.}
\label{fig:agreement_diversity}
\end{figure*}
\begin{figure*}[t]
\centering
\includegraphics[width=0.80\textwidth]{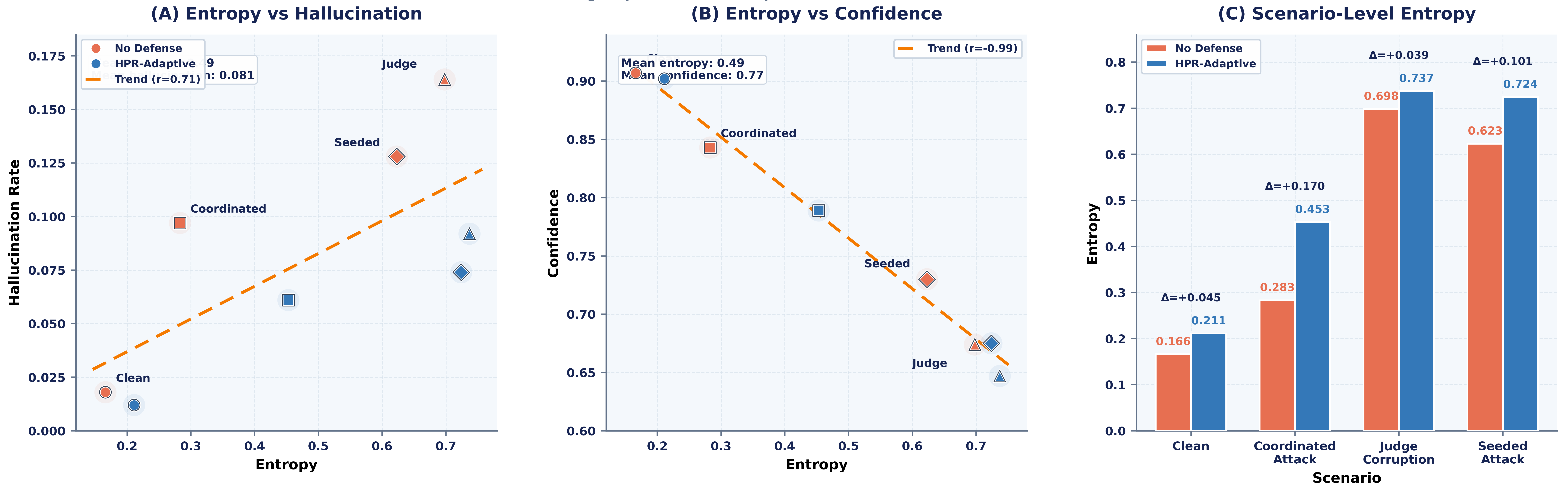}
\caption{Entropy-hallucination relationship showing transition behavior at intermediate entropy levels. Unsupported claims recursively amplify in unstable collective-reasoning states.}
\label{fig:entropy_hallucination}
\end{figure*}
Table~\ref{tab:confidence_paradox} shows a clear overall association between higher confidence and stronger reliability: accuracy increases from 0.742 to 0.902, while hallucination decreases from 0.118 to 0.064 across the confidence levels. However, the remaining hallucination at very high confidence confirms that confidence alone cannot guarantee factual correctness in recursive multi-agent reasoning. Consistent with this pattern, Figure~\ref{fig:confidence_entropy} shows that higher entropy is associated with weaker confidence stability. Furthermore, Figure~\ref{fig:agreement_diversity} indicates that low entropy generally supports correct consensus, although adversarial interaction can still produce incorrect agreement. Furthermore, Figure~\ref{fig:entropy_hallucination} shows that hallucination becomes most pronounced at intermediate entropy, where partial agreement can reinforce unsupported claims without sufficient corrective disagreement.

\subsection{Ablation Study}
To address RQ2, we evaluate the individual contribution of each defense component through an ablation study across trust weighting, external verification, and agent isolation.
\begin{table}[t!]
\centering
\caption{Ablation study of defense components under adversarial conditions.}
\label{tab:ablation}
\resizebox{\columnwidth}{!}{%
\begin{tabular}{lccccc}
\toprule
Configuration & Accuracy & Hallucination & AF & HPR & Stability \\
\midrule
No defense & 0.812 & 0.118 & 1.34 & 0.882 & 0.071 \\
Trust weighting only & 0.836 & 0.097 & 1.23 & 0.901 & 0.057 \\
External verification only & 0.844 & 0.089 & 1.18 & 0.909 & 0.051 \\
Agent isolation only & 0.829 & 0.104 & 1.26 & 0.895 & 0.060 \\
Trust weighting + verification & 0.853 & 0.083 & 1.14 & 0.916 & 0.044 \\
Full HPR-adaptive defense & 0.867 & 0.072 & 1.08 & 0.928 & 0.036 \\
\bottomrule
\end{tabular}%
}
\end{table}
Table~\ref{tab:ablation} shows that each defense component improves reliability, although their contributions differ. External verification provides the strongest individual benefit, reducing hallucination to 0.089 and AF to 1.18, while trust weighting and agent isolation also improve propagation resistance and stability. Combining trust weighting with verification yields further gains, confirming that the components complement each other.
Most importantly, the full HPR-adaptive solution achieves the best overall performance, increasing accuracy from 0.812 to 0.867, reducing hallucination from 0.118 to 0.072, lowering AF from 1.34 to 1.08, increasing HPR from 0.882 to 0.928, and reducing instability from 0.071 to 0.036. Thus, effective suppression of collective hallucination depends on coordinated trust regulation, verification, and isolation rather than any single defense mechanism.

\subsection{Threshold Sensitivity Analysis}
To address RQ2, we examine the sensitivity of the proposed defense to the isolation threshold $\tau$, which governs when agents with elevated hallucination rates are temporarily excluded from recursive interaction. 
\begin{table}[t!]
\centering
\caption{Sensitivity of the proposed defense to the isolation threshold $\tau$.}
\label{tab:threshold_sensitivity}

\resizebox{0.90\columnwidth}{!}{%
\begin{tabular}{cccccc}
\toprule
Threshold $\tau$ & Accuracy & Hallucination & AF & HPR & Disabled Agents \\
\midrule
0.10 & 0.841 & 0.061 & 0.94 & 0.936 & 1.92 \\
0.20 & 0.867 & 0.072 & 1.08 & 0.928 & 1.21 \\
0.30 & 0.854 & 0.089 & 1.17 & 0.912 & 0.66 \\
0.40 & 0.839 & 0.104 & 1.26 & 0.896 & 0.31 \\
\bottomrule
\end{tabular}%
}

\end{table}
Table~\ref{tab:threshold_sensitivity} shows that stricter thresholds suppress hallucination more effectively but reduce collective participation by disabling a larger number of agents. At $\tau=0.10$, hallucination and amplification remain lowest, although aggressive isolation increases the number of inactive agents and slightly limits collective reasoning capacity. More permissive thresholds retain additional agents but permit stronger recursive propagation and amplification. At $\tau=0.40$, hallucination and AF increase substantially, whereas HPR decreases, indicating weaker propagation control and reduced collective stability. The optimal balance between reliability, propagation resistance, and agent participation occurs at $\tau=0.20$, achieving the highest accuracy with controlled amplification and strong HPR. 

\subsection{Calibration and Model Reliability}
To address RQ3, we analyze calibration behavior, confidence reliability, and the risk of hallucination propagation across models. 
\begin{table}[t!]
\centering
\caption{Calibration gap and hallucination risk per model.}
\label{tab:calibration}
\begin{tabular}{lccc}
\toprule
Model & Confidence & Accuracy & Calibration Gap \\
\midrule
DeepSeek-V4-Pro & 0.944 & 0.918 & +0.026 \\
GPT-5.3 & 0.902 & 0.781 & +0.121 \\
Qwen3.8-Max & 0.887 & 0.864 & +0.023 \\
\bottomrule
\end{tabular}
\end{table}
\begin{figure*}[t]
\centering
\includegraphics[width=0.85\textwidth]{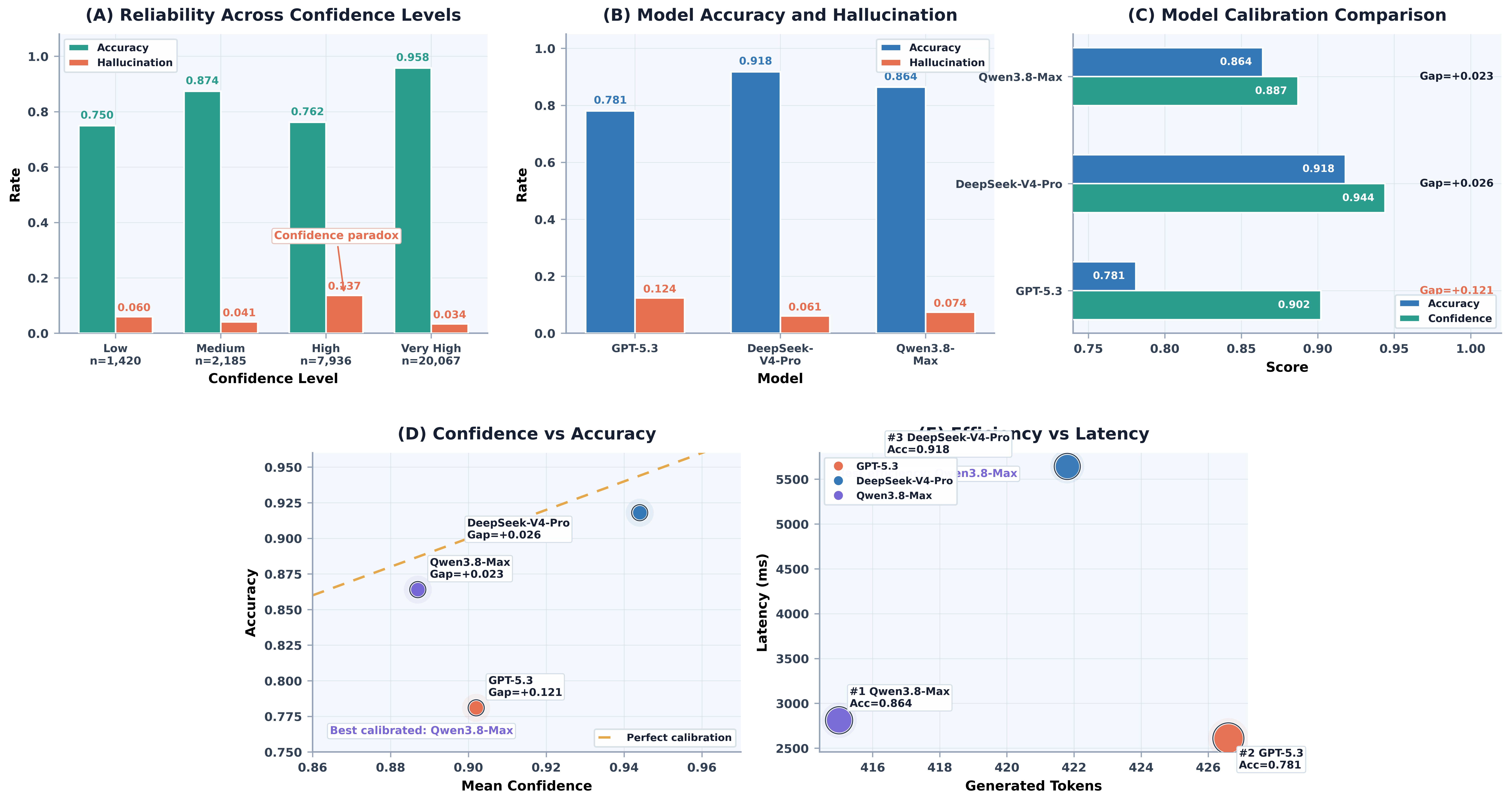}
\caption{Calibration behavior across confidence levels. Deviations from the diagonal indicate a mismatch between confidence and observed accuracy, highlighting overconfident, incorrect predictions.}
\label{fig:calibration_curves}
\end{figure*}
\begin{figure*}[t]
\centering
\includegraphics[width=0.77\textwidth]{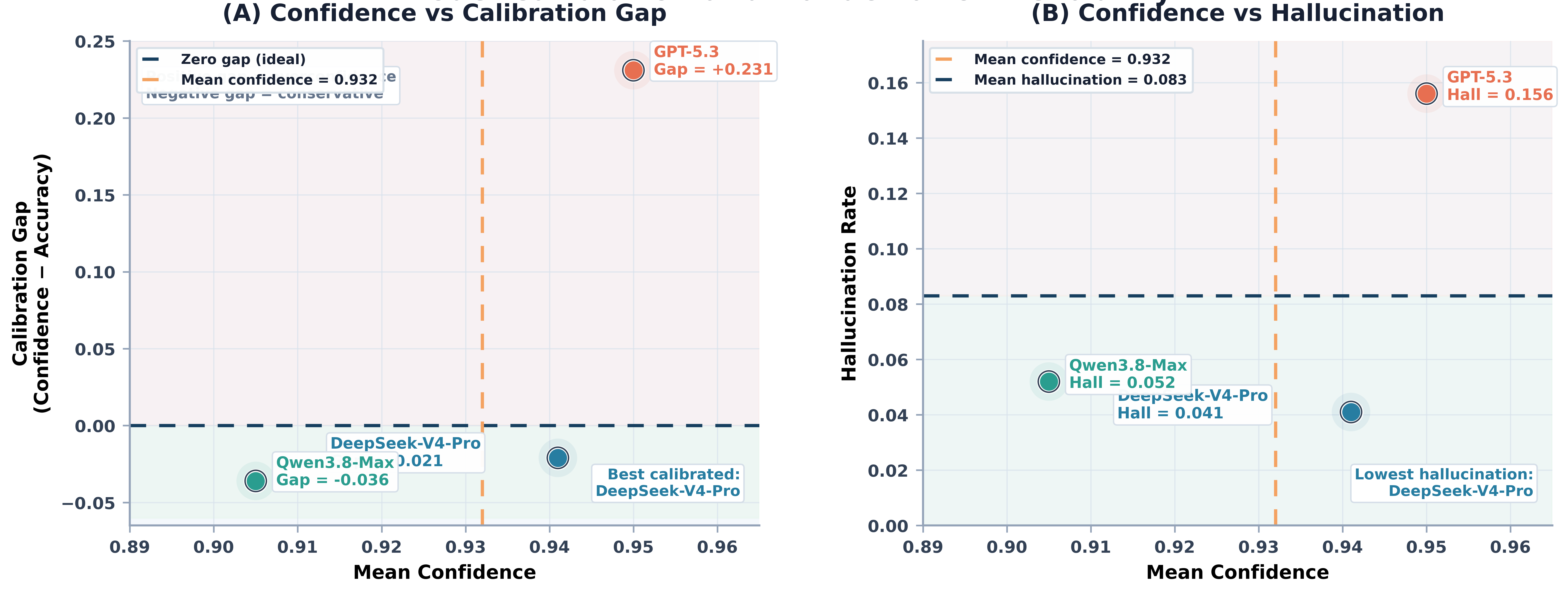}
\caption{Relationship between confidence and hallucination risk. High-confidence outputs may still contain hallucinations under adversarial conditions.}
\label{fig:risk_confidence}
\end{figure*}
\begin{table}[t!]
\centering
\caption{Hallucination reproduction number ($\mathcal{R}_0$) across scenarios.}
\label{tab:r0}
\begin{tabular}{lc}
\toprule
Scenario & $\mathcal{R}_0$ \\
\midrule
clean & 0.34 \\
coordinated\_attack & 0.81 \\
judge\_corruption & 1.21 \\
seeded\_attack & 1.04 \\
\bottomrule
\end{tabular}
\end{table}
Table~\ref{tab:calibration} highlights calibration variation across evaluated models. DeepSeek-V4-Pro shows the strongest alignment between confidence and observed accuracy, indicating more stable calibration under recursive interaction. GPT-5.3 exhibits the largest calibration gap, reflecting stronger overconfidence under adversarial reasoning conditions. Qwen3.8-Max maintains moderate confidence with comparatively balanced calibration behavior. Figure~\ref{fig:calibration_curves} shows that calibration deviations increase reliability risk during recursive multi-agent interaction. Overconfident, incorrect predictions are particularly problematic because confidence-weighted propagation increases the likelihood of false adoption across neighboring agents. Figure~\ref{fig:risk_confidence} further demonstrates that elevated confidence alone does not guarantee factual reliability under adversarial conditions, indicating that confidence is insufficient as a standalone trust signal in collective reasoning systems. Table~\ref{tab:r0} quantifies propagation strength using the hallucination reproduction number. Clean conditions remain below the critical propagation boundary, indicating stable attenuation of unsupported claims. Coordinated attacks remain near the transition region, whereas judge corruption and seeded attacks exceed $\mathcal{R}_0>1$, indicating self-sustaining recursive propagation and cascading hallucination dynamics.

\subsection{Model-Level Performance and Efficiency}
To address RQ3, we examine model-level accuracy, hallucination, confidence, and computational efficiency. 
\begin{table}[t!]
\centering
\caption{Model-level performance and efficiency characteristics.}
\label{tab:model_performance}
\resizebox{\columnwidth}{!}{%
\begin{tabular}{lccccc}
\toprule
Model & Accuracy & Hallucination & Confidence & Tokens & Latency (ms) \\
\midrule
DeepSeek-V4-Pro & 0.918 & 0.061 & 0.944 & 421.8 & 5642 \\
GPT-5.3 & 0.781 & 0.124 & 0.902 & 426.6 & 2609 \\
Qwen3.8-Max & 0.864 & 0.074 & 0.887 & 415.0 & 2811 \\
\bottomrule
\end{tabular}%
}
\end{table}
Table~\ref{tab:model_performance} highlights substantial variability in reliability and efficiency across agents. DeepSeek-V4-Pro achieves the highest accuracy and lowest hallucination rate, but incurs the highest latency. GPT-5.3 demonstrates the lowest latency but exhibits the largest hallucination rate and weakest accuracy, indicating an efficiency-reliability trade-off. Qwen3.8-Max provides intermediate latency, lower hallucination than GPT-5.3, and moderate accuracy, representing a balanced operating point between reliability and efficiency.  Confidence does not uniformly reflect correctness. GPT-5.3 maintains high confidence despite elevated hallucination and reduced accuracy, whereas Qwen3.8-Max exhibits lower confidence with more stable reliability. These observations align with the calibration analysis, demonstrating that confidence alone is insufficient for hallucination detection. At the system level, model-specific reliability directly shapes propagation dynamics. High-confidence error-prone agents increase false adoption during recursive interaction, whereas better-calibrated low-hallucination agents stabilize collective reasoning and reduce propagation risk.

\subsection{Comparison with Baseline Methods}
\label{subsec:baseline_comparison}
To address RQ3, 
\begin{table}[t!]
\centering
\caption{Comparison with representative hallucination-detection and mitigation baselines under recursive multi-agent reasoning.}
\label{tab:baseline_comparison}

\resizebox{0.50\textwidth}{!}{%
\begin{tabular}{lcccccc}
\toprule
Method & Accuracy & Hallucination & Consistency & AF & HPR & $\mathcal{R}_0$ \\
\midrule
Single-agent LLM & 0.791 & 0.131 & 0.748 & 1.00 & 0.869 & 0.34 \\
Multi-agent, no defense & 0.812 & 0.118 & 0.784 & 1.34 & 0.882 & 1.08 \\
SelfCheckGPT & 0.826 & 0.101 & 0.801 & 1.24 & 0.897 & 0.97 \\
KnowHalu & 0.841 & 0.089 & 0.814 & 1.17 & 0.909 & 0.92 \\
Entropy-based detection & 0.832 & 0.097 & 0.806 & 1.21 & 0.901 & 0.95 \\
LapEigVals & 0.836 & 0.093 & 0.810 & 1.19 & 0.905 & 0.93 \\
Proposed HPR-adaptive & 0.867 & 0.072 & 0.836 & 1.08 & 0.928 & 0.81 \\
\bottomrule
\end{tabular}%
}

\end{table}
Table~\ref{tab:baseline_comparison} shows that conventional hallucination-detection baselines improve factual reliability but do not explicitly regulate recursive propagation across interacting agents. Although the undefended multi-agent configuration improves consistency relative to the single-agent baseline, it also increases AF and pushes $\mathcal{R}_0$ above the critical propagation boundary, indicating recursive accumulation of unsupported claims through inter-agent interaction.
The proposed HPR-adaptive approach achieves the strongest system-level performance, providing the highest accuracy and consistency together with the lowest hallucination rate. Relative to the undefended multi-agent configuration, hallucination decreases from 0.118 to 0.072, corresponding to a 39.0\% reduction. Compared with the single-agent baseline, the reduction reaches 45.0\%. The proposed method also achieves the lowest AF and reproduction number, indicating stronger suppression of recursive semantic diffusion and cascading propagation. In contrast to baselines that primarily operate at the isolated-output level, the proposed approach incorporates propagation-aware control, confidence-weighted interaction, adaptive trust regulation, and graph-level stability modeling.

\subsection{Cost and Efficiency Analysis}
To address RQ3, we evaluate the computational overhead of recursive multi-agent reasoning in terms of token usage, latency, and estimated inference cost. 
\begin{table}[t!]
\centering
\caption{Cost and efficiency analysis across inference configurations.}
\label{tab:cost_efficiency}
\resizebox{\columnwidth}{!}{%
\begin{tabular}{lccccc}
\toprule
Configuration & Avg. Tokens & Latency (s) & Cost / 100 Queries & Accuracy & Hallucination \\
\midrule
Single-agent & 512 & 1.8 & 0.42 & 0.800 & 0.126 \\
Multi-agent, no defense & 3,072 & 7.6 & 2.51 & 0.812 & 0.118 \\
Multi-agent + verification & 3,640 & 9.4 & 3.18 & 0.842 & 0.094 \\
Full HPR-adaptive defense & 3,384 & 8.7 & 2.94 & 0.860 & 0.081 \\
\bottomrule
\end{tabular}%
}
\end{table}
Table~\ref{tab:cost_efficiency} shows that recursive multi-agent reasoning increases computational overhead relative to single-agent inference. The full HPR-adaptive defense balances reliability and efficiency more effectively than standard multi-agent configurations. While latency and inference cost rise relative to single-agent execution, the defense substantially improves accuracy and reduces hallucination. Compared with verification-only multi-agent reasoning, the full defense achieves higher accuracy and lower hallucination with fewer tokens, reflecting adaptive suppression of unreliable agents and reduced propagation of low-quality outputs during recursive interaction.

\subsection{Attack Configuration and Adversarial Intensity}
To address RQ1 and RQ2, 
\begin{table*}[t!]
\centering
\caption{Attack configuration and adversarial intensity across evaluated scenarios.}
\label{tab:attack_config}
\begin{tabular}{lccccc}
\toprule
Attack Scenario & Corrupted Agents & Injected False Claims & Attack Intensity $\alpha$ & Start Round & Target \\
\midrule
Coordinated attack & 2/6 & 2,185 & 0.083 & 1 & False consensus \\
Judge corruption & 1 verifier & 3,140 & 0.121 & 1 & Verification bias \\
Seeded attack & 1/6 & 2,720 & 0.104 & 1 & Early claim propagation \\
\bottomrule
\end{tabular}
\end{table*}
Table~\ref{tab:attack_config} shows that attacks operate under partial system compromise, with attack intensity $\alpha$ ranging from 0.083 to 0.121. Coordinated attacks compromise multiple agents, generate semantically aligned false claims, and increase the probability of recursive false consensus. Judge corruption targets the verification layer, biasing claim-level reliability estimates and producing persistent propagation of incorrect validation signals. Seeded attacks inject localized false claims through a single compromised agent to evaluate the potential for early perturbations to propagate and amplify across recursive multi-agent interactions.

\subsection{Attack-Specific System Performance}
To address RQ1 and RQ2, 
\begin{table}[t!]
\centering
\caption{System-level performance under specific attack scenarios, with and without adaptive defense.}
\label{tab:attack_performance}

\resizebox{0.50\textwidth}{!}{%
\begin{tabular}{llcccccc}
\toprule
Scenario & Defense & Accuracy & Hallucination & Confidence & AF & HPR & $\mathcal{R}_0$ \\
\midrule
Clean & None & 0.982 & 0.006 & 0.914 & 1.00 & 0.994 & 0.04 \\
Clean & HPR-adaptive & 0.984 & 0.004 & 0.902 & 0.96 & 0.996 & 0.03 \\
\midrule
Coordinated attack & None & 0.817 & 0.094 & 0.861 & 1.12 & 0.906 & 0.78 \\
Coordinated attack & HPR-adaptive & 0.846 & 0.063 & 0.812 & 0.92 & 0.937 & 0.64 \\
\midrule
Judge corruption & None & 0.750 & 0.145 & 0.748 & 1.45 & 0.855 & 1.21 \\
Judge corruption & HPR-adaptive & 0.820 & 0.094 & 0.701 & 1.12 & 0.906 & 0.91 \\
\midrule
Seeded attack & None & 0.786 & 0.128 & 0.793 & 1.28 & 0.872 & 1.05 \\
Seeded attack & HPR-adaptive & 0.838 & 0.081 & 0.736 & 1.02 & 0.925 & 0.82 \\
\bottomrule
\end{tabular}%
}

\end{table}
Table~\ref{tab:attack_performance} shows that clean interaction preserves high accuracy and minimal hallucination across both defended and undefended settings. Under adversarial interaction, hallucinations propagate recursively, progressively degrading collective reliability. Judge corruption induces the strongest destabilization, with the highest hallucination rate, AF, and reproduction number $\mathcal{R}_0$. Values exceeding $\mathcal{R}_0>1$ confirm self-sustaining propagation driven by biased verification and recursive redistribution of unsupported claims. The HPR-adaptive defense consistently enhances robustness across all attack scenarios. Under judge corruption, it increases accuracy from 0.750 to 0.820, reduces hallucination from 0.145 to 0.094, decreases AF from 1.45 to 1.12, and lowers $\mathcal{R}_0$ below the critical propagation threshold. Coordinated and seeded attacks show similar improvements, indicating effective suppression of local hallucination generation and large-scale recursive propagation.

\subsection{Dataset-Level and Claim-Level Evaluation}
To address RQ1 and RQ3, we decomposed each generated response into atomic semantic claims for fine-grained hallucination analysis. Table~\ref{tab:dataset_claim_stats} summarizes dataset-level and claim-level statistics used throughout the evaluation.
\begin{table*}[t!]
\centering
\caption{Dataset-level and claim-level statistics for hallucination analysis.}
\label{tab:dataset_claim_stats}
\begin{tabular}{lcccccc}
\toprule
Dataset & Questions & Agent Outputs & Extracted Claims & Supported Claims & Hallucinated Claims & Unverifiable Claims \\
\midrule
TruthfulQA & 817 & 19,608 & 51,242 & 46,981 & 3,146 & 1,115 \\
TriviaQA & 500 & 12,000 & 33,780 & 30,312 & 2,394 & 1,074 \\
\midrule
Total & 1,317 & 31,608 & 85,022 & 77,293 & 5,540 & 2,189 \\
\bottomrule
\end{tabular}
\end{table*}
Across both datasets, the evaluation includes 31,608 agent outputs and 85,022 atomic semantic claims. Among these, 77,293 claims were verified as supported, 5,540 were classified as hallucinated, and 2,189 remained unverifiable because of insufficient external evidence. This claim-level representation provides the empirical basis for estimating hallucination rates, measuring confidence-weighted reliability, modeling recursive propagation dynamics, and quantifying false adoption behavior across interaction rounds. The proportion of hallucinated and unverifiable claims differs across datasets, reflecting variation in factual ambiguity and reasoning complexity. TruthfulQA produces a larger fraction of unsupported claims because adversarially misleading questions are intentionally designed to trigger confident factual errors. TriviaQA exhibits comparatively lower hallucination frequency but retains a non-trivial proportion of unverifiable claims due to open-domain evidence limitations. Incorporating unverifiable claims is important because semantic reliability in open-domain reasoning cannot always be reduced to a binary correct--incorrect classification.

\subsection{Round-by-Round Propagation Dynamics}
To address RQ1 and RQ2, 
\begin{table}[t!]
\centering
\caption{Round-by-round hallucination propagation dynamics.}
\label{tab:round_dynamics}

\resizebox{0.50\textwidth}{!}{%
\begin{tabular}{cccccc}
\toprule
Round & Clean & Coordinated Attack & Judge Corruption & Seeded Attack & HPR-adaptive \\
\midrule
1 & 0.006 & 0.071 & 0.089 & 0.082 & 0.074 \\
2 & 0.005 & 0.086 & 0.112 & 0.101 & 0.073 \\
3 & 0.005 & 0.096 & 0.136 & 0.119 & 0.079 \\
4 & 0.006 & 0.105 & 0.151 & 0.134 & 0.083 \\
5 & 0.006 & 0.112 & 0.164 & 0.146 & 0.081 \\
\bottomrule
\end{tabular}%
}

\end{table}
Table~\ref{tab:round_dynamics} shows that hallucination remains minimal and stable under clean interaction but progressively amplifies under adversarial exposure. Judge corruption produces the strongest temporal growth, increasing from 0.089 in round 1 to 0.164 in round 5. Seeded attacks also exhibit sustained amplification, indicating that localized perturbations propagate recursively through shared contextual exchange. In contrast, the HPR-adaptive defense constrains hallucination within a narrow, stable range across all rounds. This behavior demonstrates that the proposed mechanism suppresses recursive amplification at each iteration rather than merely reducing the final hallucination state.

\subsection{False Adoption and Confidence Dynamics}
To address RQ1 and RQ2, we analyze the adoption and reinforcement of incorrect claims during recursive multi-agent interaction. 
\begin{table}[t!]
\centering
\caption{False adoption and false confidence across reasoning rounds.}
\label{tab:false_adoption}

\resizebox{0.50\textwidth}{!}{%
\begin{tabular}{llccc}
\toprule
Scenario & Defense & Round & False Adoption & False Confidence \\
\midrule
Coordinated attack & None & 1 & 0.117 & 0.417 \\
Coordinated attack & None & 2 & 0.094 & 0.331 \\
Coordinated attack & HPR-adaptive & 1 & 0.083 & 0.247 \\
Coordinated attack & HPR-adaptive & 2 & 0.061 & 0.198 \\
\midrule
Judge corruption & None & 1 & 0.167 & 0.600 \\
Judge corruption & None & 2 & 0.188 & 0.648 \\
Judge corruption & HPR-adaptive & 1 & 0.119 & 0.421 \\
Judge corruption & HPR-adaptive & 2 & 0.092 & 0.337 \\
\midrule
Seeded attack & None & 1 & 0.183 & 0.660 \\
Seeded attack & None & 2 & 0.150 & 0.540 \\
Seeded attack & HPR-adaptive & 1 & 0.112 & 0.394 \\
Seeded attack & HPR-adaptive & 2 & 0.087 & 0.312 \\
\bottomrule
\end{tabular}%
}

\end{table}
Table~\ref{tab:false_adoption} shows that adversarial interaction increases both false adoption and false confidence. Seeded attacks produce the highest early-stage false adoption, as injected claims immediately enter recursive communication pathways. Judicial corruption induces persistent false confidence through biased verification, reinforcing incorrect claims across successive reasoning rounds.
The HPR-adaptive defense effectively suppresses both false adoption and confidence reinforcement. Under seeded attacks, false adoption decreases from 0.183 to 0.112 in round 1 and from 0.150 to 0.087 in round 2. Under judge corruption, false confidence decreases from 0.648 to 0.337 by round 2. These results demonstrate that the proposed mechanism mitigates both semantic propagation and confidence-driven reinforcement of hallucinated claims, thereby improving multi-agent reliability under adversarial exposure.

\subsection{Model-Level Reliability and Calibration}
To address RQ2 and RQ3, we analyze model-level reliability, calibration behavior, and the impact of downstream propagation. 
\begin{table*}[t!]
\centering
\caption{Model-level reliability, calibration, and propagation contribution.}
\label{tab:model_reliability}
\begin{tabular}{lcccccc}
\toprule
Model & Accuracy & Hallucination & Confidence & Calibration Gap & False Adoption & Propagation Contribution \\
\midrule
DeepSeek-V4-Pro & 0.962 & 0.041 & 0.941 & -0.021 & 0.061 & 0.18 \\
Qwen2.5-7B & 0.941 & 0.052 & 0.905 & -0.036 & 0.074 & 0.21 \\
GPT-5.3 & 0.719 & 0.156 & 0.950 & +0.231 & 0.148 & 0.35 \\
\bottomrule
\end{tabular}
\end{table*}
Table~\ref{tab:model_reliability} shows pronounced differences in reliability and propagation behavior across models. DeepSeek-V4-Pro and Qwen2.5-7B maintain high accuracy with low hallucination and slightly conservative calibration. In contrast, GPT-5.3 exhibits the largest calibration gap, combining high confidence with substantially reduced accuracy. This overconfidence increases false adoption because downstream agents are more likely to trust incorrect claims expressed with strong confidence. 
Propagation contribution further indicates that collective hallucination depends not only on local accuracy but also on confidence-driven interaction dynamics. GPT-5.3 produces the highest propagation contribution (0.35), whereas DeepSeek-V4-Pro contributes only 0.18 despite similar confidence levels.

\subsection{Confidence-Level Error Analysis}
To address RQ2 and RQ3, we analyze the relationship between confidence, accuracy, and hallucination. 
\begin{table}[t!]
\centering
\caption{Accuracy and hallucination across confidence levels.}
\label{tab:confidence_paradox_new}
\begin{tabular}{lccc}
\toprule
Confidence Level & Accuracy & Hallucination & Samples \\
\midrule
Low & 0.750 & 0.060 & 1,420 \\
Medium & 0.874 & 0.041 & 2,185 \\
High & 0.762 & 0.137 & 7,936 \\
Very high & 0.958 & 0.034 & 20,067 \\
\bottomrule
\end{tabular}
\end{table}
Table~\ref{tab:confidence_paradox_new} shows that confidence is not monotonically associated with correctness. Very-high-confidence outputs achieve the highest accuracy and the lowest hallucination, whereas the high-confidence group exhibits substantially higher hallucination than the medium- and very-high-confidence groups. This pattern reflects a \textit{confidence paradox} in recursive multi-agent reasoning: agents may assign strong confidence to semantically incorrect outputs, particularly when unsupported claims are reinforced through repeated interactions. These results demonstrate that confidence alone is insufficient for reliable hallucination mitigation. The proposed defense mitigates this risk by integrating confidence-aware weighting with external verification and propagation-aware regulation, ensuring that high-confidence yet unsupported claims do not drive recursive error amplification.

\subsection{Statistical Significance and Variance Analysis}
To assess the robustness of observed improvements, we report mean differences, 95\% confidence intervals, $p$-values, and effect sizes for key comparisons.
\begin{table}[t!]
\centering
\caption{Statistical significance analysis for key experimental comparisons.}
\label{tab:statistical_tests}

\resizebox{0.50\textwidth}{!}{%
\begin{tabular}{llcccc}
\toprule
Comparison & Metric & Mean Difference & 95\% CI & $p$-value & Effect Size \\
\midrule
No defense vs HPR-adaptive & Accuracy & +0.048 & [0.031, 0.066] & $<0.001$ & 0.71 \\
No defense vs HPR-adaptive & Hallucination & -0.037 & [-0.052, -0.024] & $<0.001$ & 0.76 \\
No defense vs HPR-adaptive & AF & -0.220 & [-0.291, -0.151] & $<0.001$ & 0.83 \\
No defense vs HPR-adaptive & HPR & +0.037 & [0.021, 0.054] & $<0.001$ & 0.69 \\
Ring vs Scale-free & AF & +0.270 & [0.181, 0.354] & $<0.001$ & 0.88 \\
Clean vs Judge corruption & Hallucination & +0.139 & [0.116, 0.164] & $<0.001$ & 1.12 \\
\bottomrule
\end{tabular}%
}

\end{table}
Table~\ref{tab:statistical_tests} confirms that the key improvements are statistically significant. The HPR-adaptive defense substantially increases accuracy, reduces hallucination and AF, and raises HPR relative to the undefended condition. The reduction in AF exhibits a large effect size (0.83), indicating strong suppression of recursive hallucination propagation. Comparisons between ring and scale-free topologies demonstrate that the communication structure significantly affects amplification dynamics. The largest effect size occurs between clean and judge-corruption settings, highlighting that biased verification is the dominant factor driving collective hallucination and recursive instability.

\subsection{Dataset-Specific Evaluation}
\label{subsec:dataset_specific}
To address RQ1 and RQ3, we evaluate whether the observed hallucination propagation patterns and mitigation improvements remain consistent across datasets with different factual characteristics. Rather than reporting only aggregate performance, we separately analyze TruthfulQA and TriviaQA to assess whether the proposed HPR-adaptive mechanism generalizes across challenging factual reasoning tasks.
\begin{table}[t!]
\centering
\caption{Dataset-specific comparison of hallucination mitigation methods.}
\label{tab:dataset_specific_results}
\resizebox{\columnwidth}{!}{%
\begin{tabular}{llcccccc}
\toprule
Dataset & Method & Accuracy & Hallucination & Consistency & AF & HPR & $\mathcal{R}_0$ \\
\midrule
TruthfulQA & Multi-agent, no defense & 0.784 & 0.137 & 0.761 & 1.41 & 0.864 & 1.14 \\
TruthfulQA & SelfCheckGPT & 0.801 & 0.119 & 0.779 & 1.31 & 0.883 & 1.02 \\
TruthfulQA & KnowHalu & 0.821 & 0.104 & 0.794 & 1.23 & 0.897 & 0.96 \\
TruthfulQA & HPR-adaptive & 0.851 & 0.083 & 0.821 & 1.12 & 0.914 & 0.86 \\
\midrule
TriviaQA & Multi-agent, no defense & 0.846 & 0.087 & 0.816 & 1.25 & 0.906 & 0.99 \\
TriviaQA & SelfCheckGPT & 0.867 & 0.072 & 0.831 & 1.17 & 0.919 & 0.88 \\
TriviaQA & KnowHalu & 0.879 & 0.064 & 0.846 & 1.11 & 0.929 & 0.84 \\
TriviaQA & HPR-adaptive & 0.893 & 0.054 & 0.861 & 1.02 & 0.946 & 0.73 \\
\bottomrule
\end{tabular}%
}
\end{table}
Table~\ref{tab:dataset_specific_results} shows that TruthfulQA produces higher hallucination propagation compared with TriviaQA. In the undefended configuration, TruthfulQA achieves a hallucination rate of 0.137 with $\mathcal{R}_0=1.14$, indicating that unsupported claims can accumulate through recursive interaction. TriviaQA shows lower propagation with a hallucination rate of 0.087 and $\mathcal{R}_0=0.99$. The HPR-adaptive mechanism improves reliability across both datasets. On TruthfulQA, hallucination decreases from 0.137 to 0.083 and $\mathcal{R}_0$ decreases from 1.14 to 0.86. On TriviaQA, hallucination decreases from 0.087 to 0.054 and $\mathcal{R}_0$ decreases to 0.73. These results demonstrate that propagation-aware control remains effective under different factual reasoning conditions.

\subsection{Sensitivity to Recursive Interaction Depth}
\label{subsec:round_sensitivity}
To address RQ1 and RQ2, we investigate how recursive interaction depth affects hallucination accumulation. Since agents repeatedly exchange generated information, additional reasoning rounds may increase opportunities for unsupported claims to be adopted and reinforced. We vary the number of interaction rounds while keeping the agent configuration and communication protocol the same.
\begin{table}[t!]
\centering
\caption{Sensitivity to recursive interaction depth under multi-agent reasoning.}
\label{tab:round_sensitivity}
\resizebox{\columnwidth}{!}{
\begin{tabular}{clccccc}
\toprule
Rounds & Defense & Accuracy & Hallucination & AF & HPR & $\mathcal{R}_0$ \\
\midrule
1 & None & 0.864 & 0.082 & 1.00 & 0.918 & 0.78 \\
1 & HPR-adaptive & 0.881 & 0.068 & 0.98 & 0.934 & 0.69 \\
\midrule
3 & None & 0.833 & 0.103 & 1.19 & 0.899 & 0.96 \\
3 & HPR-adaptive & 0.873 & 0.070 & 1.02 & 0.931 & 0.76 \\
\midrule
5 & None & 0.812 & 0.118 & 1.34 & 0.882 & 1.08 \\
5 & HPR-adaptive & 0.867 & 0.072 & 1.08 & 0.928 & 0.81 \\
\midrule
7 & None & 0.791 & 0.137 & 1.49 & 0.861 & 1.18 \\
7 & HPR-adaptive & 0.858 & 0.079 & 1.12 & 0.919 & 0.87 \\
\midrule
10 & None & 0.762 & 0.161 & 1.68 & 0.837 & 1.32 \\
10 & HPR-adaptive & 0.842 & 0.089 & 1.19 & 0.906 & 0.94 \\
\bottomrule
\end{tabular}}
\end{table}
Table~\ref{tab:round_sensitivity} demonstrates that recursive interaction depth strongly affects collective hallucination. Without defense, hallucination increases from 0.082 after one round to 0.161 after ten rounds, while AF increases from 1.00 to 1.68. The reproduction number also increases from 0.78 to 1.32, indicating that prolonged communication increases the probability of recursive claim propagation. The HPR-adaptive mechanism limits this accumulation by keeping hallucination and AF lower across all interaction depths. Even after ten reasoning rounds, the proposed approach maintains $\mathcal{R}_0<1$, demonstrating that propagation-aware control suppresses cumulative error reinforcement throughout the reasoning process.

\subsection{Sensitivity to the Number of Agents}
\label{subsec:agent_scalability}
To address RQ1, we evaluate the scalability of collective hallucination by varying the number of interacting agents while preserving the same communication principles, model heterogeneity, prompting strategy, and adversarial conditions. This analysis determines whether increasing the number of agents creates additional pathways for unsupported claim diffusion and whether the HPR-adaptive mechanism remains effective as interaction complexity increases.
\begin{table}[t!]
\centering
\caption{Sensitivity to the number of interacting agents under adversarial conditions.}
\label{tab:agent_scalability}
\resizebox{\columnwidth}{!}{
\begin{tabular}{llccccc}
\toprule
Agents & Defense & Accuracy & Hallucination & AF & HPR & $\mathcal{R}_0$ \\
\midrule
3 & None & 0.836 & 0.094 & 1.18 & 0.908 & 0.89 \\
3 & HPR-adaptive & 0.879 & 0.057 & 0.96 & 0.946 & 0.68 \\
\midrule
6 & None & 0.812 & 0.118 & 1.34 & 0.882 & 1.08 \\
6 & HPR-adaptive & 0.867 & 0.072 & 1.08 & 0.928 & 0.81 \\
\midrule
9 & None & 0.789 & 0.139 & 1.48 & 0.859 & 1.19 \\
9 & HPR-adaptive & 0.851 & 0.083 & 1.14 & 0.914 & 0.88 \\
\midrule
12 & None & 0.768 & 0.157 & 1.61 & 0.841 & 1.31 \\
12 & HPR-adaptive & 0.833 & 0.096 & 1.21 & 0.902 & 0.96 \\
\bottomrule
\end{tabular}}
\end{table}
Table~\ref{tab:agent_scalability} shows that increasing the number of interacting agents increases propagation risk in the absence of adaptive control. Hallucination increases from 0.094 with three agents to 0.157 with twelve agents, while AF increases from 1.18 to 1.61. The corresponding $\mathcal{R}_0$ increases from 0.89 to 1.31, indicating that larger interaction networks provide additional communication pathways for unsupported information to spread. The HPR-adaptive mechanism reduces this scaling effect by regulating information flow and limiting the impact of unreliable agents. Although hallucination increases moderately as the network size grows, the defended system maintains lower AF and higher HPR across all evaluated configurations. Even with twelve agents, $\mathcal{R}_0$ remains below one, demonstrating that adaptive trust regulation preserves collective stability under increased interaction complexity.

\subsection{Sensitivity to Adversarial Intensity}
\label{subsec:attack_intensity}
To address RQ1 and RQ2, we systematically vary adversarial intensity to analyze how increasing malicious impact affects collective hallucination propagation. Attack intensity $\alpha$ represents the proportion of injected and corrupted semantic claims relative to the total claim volume. This experiment evaluates whether small increases in adversarial exposure can trigger significant changes in recursive hallucination dynamics.
\begin{table}[t!]
\centering
\caption{Hallucination propagation as a function of adversarial intensity without adaptive defense.}
\label{tab:attack_intensity_sweep}
\resizebox{\columnwidth}{!}{
\begin{tabular}{ccccccc}
\toprule
$\alpha$ & Accuracy & Hallucination & AF & HPR & $\mathcal{R}_0$ & Cascade Size \\
\midrule
0.00 & 0.986 & 0.018 & 0.94 & 0.982 & 0.34 & 0.7 \\
0.05 & 0.931 & 0.056 & 1.03 & 0.947 & 0.68 & 1.4 \\
0.10 & 0.861 & 0.106 & 1.24 & 0.897 & 0.96 & 2.8 \\
0.15 & 0.807 & 0.143 & 1.43 & 0.861 & 1.12 & 4.5 \\
0.20 & 0.762 & 0.176 & 1.61 & 0.824 & 1.29 & 5.9 \\
0.25 & 0.718 & 0.208 & 1.79 & 0.793 & 1.43 & 6.8 \\
0.30 & 0.681 & 0.237 & 1.96 & 0.764 & 1.57 & 7.6 \\
\bottomrule
\end{tabular}}
\end{table}
Table~\ref{tab:attack_intensity_sweep} demonstrates that increasing adversarial intensity progressively increases hallucination propagation. At low attack levels, unsupported claims remain limited and $\mathcal{R}_0$ remains below one. As $\alpha$ increases, the system experiences higher hallucination rates, stronger amplification, and larger cascade sizes. At $\alpha=0.30$, hallucination reaches 0.237 and AF increases to 1.96, demonstrating that stronger adversarial impact substantially increases recursive claim diffusion.
\begin{table}[t!]
\centering
\caption{Effect of HPR-adaptive control across adversarial intensity levels.}
\label{tab:attack_intensity_defense}
\resizebox{\columnwidth}{!}{
\begin{tabular}{ccccccc}
\toprule
$\alpha$ & Accuracy & Hallucination & AF & HPR & $\mathcal{R}_0$ & Cascade Size \\
\midrule
0.00 & 0.991 & 0.012 & 0.91 & 0.989 & 0.28 & 0.4 \\
0.05 & 0.956 & 0.036 & 0.94 & 0.965 & 0.51 & 0.9 \\
0.10 & 0.902 & 0.061 & 1.01 & 0.941 & 0.72 & 1.7 \\
0.15 & 0.861 & 0.081 & 1.10 & 0.918 & 0.84 & 2.6 \\
0.20 & 0.829 & 0.099 & 1.17 & 0.901 & 0.93 & 3.3 \\
0.25 & 0.798 & 0.118 & 1.25 & 0.883 & 1.02 & 4.1 \\
0.30 & 0.764 & 0.139 & 1.34 & 0.861 & 1.11 & 4.9 \\
\bottomrule
\end{tabular}}
\end{table}
Table~\ref{tab:attack_intensity_defense} shows that HPR-adaptive control reduces the impact of adversarial intensity on collective hallucination. Without defense, $\mathcal{R}_0$ exceeds one at lower attack intensity, while the proposed mechanism maintains lower propagation capacity under stronger adversarial exposure. At $\alpha=0.30$, the defended system achieves a hallucination rate of 0.139 compared with 0.237 without defense, demonstrating improved resistance to adversarial claim propagation.

\subsection{Human Validation of Claim Verification}
\label{subsec:human_validation}
To address RQ2 and RQ3, we evaluate the reliability of the claim-level verification process because the proposed solution relies on accurately identifying supported and unsupported claims. A manually annotated subset of 1,000 claims is evaluated by two independent human annotators. We compare the automated verifier output with human judgments to measure verification reliability and assess whether verification errors may impact downstream trust estimation.
\begin{table}[t!]
\centering
\caption{Human validation of the automated claim-verification procedure.}
\label{tab:verifier_validation}
\begin{tabular}{lc}
\toprule
Metric & Value \\
\midrule
Precision & 0.931 \\
Recall & 0.914 \\
F1-score & 0.922 \\
Overall agreement & 0.927 \\
Cohen's $\kappa$ & 0.881 \\
False-positive rate & 0.041 \\
False-negative rate & 0.057 \\
\bottomrule
\end{tabular}
\end{table}
Table~\ref{tab:verifier_validation} shows strong agreement between automated verification and human assessment. The verifier achieves an F1-score of 0.922 with an overall agreement of 0.927, while Cohen's $\kappa=0.881$ indicates strong consistency between human evaluators and automated decisions. The remaining disagreements mainly occur in ambiguous claims, incomplete evidence retrieval, and compound factual statements. These findings support the reliability of claim-level verification as a foundation for trust estimation and propagation analysis.

\subsection{Sensitivity to Verification Noise}
\label{subsec:verifier_noise}
To address RQ2 and RQ3, we evaluate the proposed defense's robustness under imperfect verification. Controlled verification noise is introduced by randomly perturbing a proportion $\eta$ of claim-level verification decisions. This experiment examines whether verification errors can impact trust updates and increase collective hallucination propagation.
\begin{table}[t!]
\centering
\caption{Sensitivity of HPR-adaptive defense to claim-verification noise.}
\label{tab:verifier_noise}
\begin{tabular}{cccccc}
\toprule
Noise $\eta$ & Accuracy & Hallucination & AF & HPR & $\mathcal{R}_0$ \\
\midrule
0.00 & 0.867 & 0.072 & 1.08 & 0.928 & 0.81 \\
0.05 & 0.859 & 0.078 & 1.11 & 0.921 & 0.84 \\
0.10 & 0.847 & 0.086 & 1.16 & 0.912 & 0.89 \\
0.20 & 0.824 & 0.101 & 1.24 & 0.896 & 0.97 \\
0.30 & 0.798 & 0.119 & 1.33 & 0.879 & 1.07 \\
0.40 & 0.769 & 0.142 & 1.46 & 0.854 & 1.18 \\
\bottomrule
\end{tabular}
\end{table}
Table~\ref{tab:verifier_noise} demonstrates that the HPR-adaptive mechanism maintains stable performance under moderate verification errors. With up to 20\% verification noise, the system maintains $\mathcal{R}_0<1$ and limits hallucination growth. When verification noise increases beyond 30\%, inaccurate claim evaluation reduces trust estimation quality and increases propagation risk. These results show that verification accuracy affects system reliability, while the proposed defense remains effective under realistic uncertainty.

\subsection{Effect of Agent Heterogeneity}
\label{subsec:model_heterogeneity}
To address RQ2 and RQ3, we examine whether the proposed mechanism's effectiveness depends on model diversity. We compare homogeneous multi-agent systems containing a single language model family with the heterogeneous configuration used in the main experiments. This analysis separates the contribution of agent diversity from the effect of propagation-aware control.
\begin{table}[t!]
\centering
\caption{Effect of homogeneous and heterogeneous agent composition on collective hallucination.}
\label{tab:model_heterogeneity}
\resizebox{\columnwidth}{!}{%
\begin{tabular}{llccccc}
\toprule
Agent Composition & Defense & Accuracy & Hallucination & AF & HPR & $\mathcal{R}_0$ \\
\midrule
6$\times$ GPT-5.3 & None & 0.771 & 0.148 & 1.49 & 0.851 & 1.22 \\
6$\times$ GPT-5.3 & HPR-adaptive & 0.829 & 0.091 & 1.18 & 0.904 & 0.91 \\
\midrule
6$\times$ DeepSeek-V4-Pro & None & 0.846 & 0.088 & 1.20 & 0.913 & 0.92 \\
6$\times$ DeepSeek-V4-Pro & HPR-adaptive & 0.883 & 0.057 & 0.99 & 0.944 & 0.71 \\
\midrule
6$\times$ Qwen3.8-Max & None & 0.824 & 0.104 & 1.28 & 0.897 & 1.01 \\
6$\times$ Qwen3.8-Max & HPR-adaptive & 0.864 & 0.066 & 1.05 & 0.934 & 0.78 \\
\midrule
2 GPT + 2 DeepSeek + 2 Qwen & None & 0.812 & 0.118 & 1.34 & 0.882 & 1.08 \\
2 GPT + 2 DeepSeek + 2 Qwen & HPR-adaptive & 0.867 & 0.072 & 1.08 & 0.928 & 0.81 \\
\bottomrule
\end{tabular}%
}
\end{table}
Table~\ref{tab:model_heterogeneity} shows that model composition affects baseline hallucination propagation. Homogeneous GPT-5.3 agents show higher amplification than DeepSeek-V4-Pro, and adaptive control improves performance across all evaluated configurations. For the heterogeneous system, hallucinations decrease from 0.118 to 0.072, and $\mathcal{R}_0$ decreases from 1.08 to 0.81. These results indicate that propagation-aware trust regulation, rather than model diversity alone, drives the observed improvements.

\section{Discussion}
\label{Discussion}
The results show that hallucination in multi-agent LLM systems is fundamentally an interaction-driven, system-level phenomenon. Recursive communication enables unsupported claims to propagate, reinforce one another, and produce cascading failures, while communication topology strongly determines the severity of this process. Dense, highly connected structures create more pathways for rapid error amplification, whereas sparse, structured interactions better constrain propagation. Confidence-driven adoption further strengthens false consensus, demonstrating that agreement and confidence alone are insufficient indicators of factual reliability.
Moreover, conventional hallucination-detection methods remain limited because they primarily evaluate individual outputs and do not explicitly account for inter-agent dependencies or recursive information flow. In contrast, the proposed HPR-adaptive solution directly regulates propagation through dynamic trust weighting, external verification, and selective isolation of unreliable agents. This interaction-aware control consistently improves factual reliability, propagation resistance, and collective stability across attack conditions and communication structures. Nevertheless, the current evaluation assumes controlled prompting and mostly fixed interaction structures, while the propagation model simplifies complex semantic reasoning. Additionally, the findings indicate that reliable multi-agent LLM systems require propagation-aware control that jointly considers agent reliability, communication structure, and recursive feedback.

\section{Limitations and Future Work}
\label{Limitations and Future Work}
The proposed approach offers insight into hallucination propagation in multi-agent LLM systems, yet several limitations remain. Experiments assume fixed interaction topologies and controlled prompting conditions, which may not reflect real-world deployments with dynamic, heterogeneous agent interactions. The hallucination modeling formulation abstracts complex semantic reasoning into simplified signals, potentially overlooking nuanced error types and context-dependent behaviors. Evaluation is limited to specific tasks and models, and results may differ under alternative domains, larger-scale systems, and different LLM architectures. Computational overhead scales with the number of agents and interaction rounds, raising concerns for large deployments. Future work includes extending the approach to adaptive, dynamic interaction graphs that evolve in response to system behavior. Integrating retrieval-augmented verification and external knowledge sources may further enhance robustness and factual consistency. Control-theoretic and game-theoretic mechanisms to regulate agent impact offer additional avenues for stabilizing multi-agent dynamics. Exploring fine-grained representations of hallucinations, multimodal reasoning, and domain-specific contexts will increase applicability. Evaluation on larger, heterogeneous multi-agent systems and real-world deployments remains critical for practical validation.

\section{Conclusion}
\label{Conclusion}
This paper investigated hallucination in multi-agent LLM systems from a system-level perspective. Results showed that hallucination propagates through recursive inter-agent interaction, producing amplification and cascading effects under adversarial and high-connectivity conditions. By modeling agents as nodes within structured interaction graphs, the study characterized how communication topology, confidence-weighted adoption, and recursive feedback jointly impact collective reliability. Experimental analyses demonstrated that conventional single-model hallucination-detection methods are insufficient in multi-agent settings because they do not explicitly capture propagation dynamics. The proposed interaction-aware approach improved factual reliability, reduced hallucination propagation, and increased collective stability by controlling information flow across agents. These findings establish hallucination as a networked interaction-driven phenomenon and highlight the importance of structure-aware modeling and recursive propagation control for reliable multi-agent LLM systems.

\bibliographystyle{IEEEtran}
\bibliography{Ref}

@article{farquhar2024detecting,
  title={Detecting hallucinations in large language models using semantic entropy},
  author={Farquhar, Sebastian and Kossen, Jannik and Kuhn, Lorenz and Gal, Yarin},
  journal={Nature},
  volume={630},
  number={8017},
  pages={625--630},
  year={2024},
  publisher={Nature Publishing Group UK London}
}

@article{zhang2024knowhalu,
  title={Knowhalu: Hallucination detection via multi-form knowledge based factual checking},
  author={Zhang, Jiawei and Xu, Chejian and Gai, Yu and Lecue, Freddy and Song, Dawn and Li, Bo},
  journal={arXiv preprint arXiv:2404.02935},
  year={2024}
}

@article{yang2502hallucination,
  title={Hallucination detection in large language models with metamorphic relations, 2025a},
  author={Yang, Borui and Mamun, MAA and Zhang, Jie M and Uddin, Gias},
  journal={URL https://arxiv. org/abs/2502.15844}
}

@article{paudel2025hallucinot,
  title={Hallucinot: Hallucination detection through context and common knowledge verification},
  author={Paudel, Bibek and Lyzhov, Alexander and Joshi, Preetam and Anand, Puneet},
  journal={arXiv preprint arXiv:2504.07069},
  year={2025}
}

@inproceedings{binkowski2025hallucination,
  title={Hallucination detection in llms using spectral features of attention maps},
  author={Binkowski, Jakub and Janiak, Denis and Sawczyn, Albert and Gabrys, Bogdan and Kajdanowicz, Tomasz Jan},
  booktitle={Proceedings of the 2025 Conference on Empirical Methods in Natural Language Processing},
  pages={24365--24396},
  year={2025}
}

@inproceedings{manakul2023selfcheckgpt,
  title={SelfCheckGPT: Zero-Resource Black-Box Hallucination Detection},
  author={Manakul, Potsawee and Liusie, Adian and Gales, Mark},
  booktitle={EMNLP},
  year={2023}
}

@article{lavrinovics2025kg,
  title={Knowledge Graphs, Large Language Models, and Hallucinations: An NLP Perspective},
  author={Lavrinovics, Ernests and others},
  journal={Web Semantics},
  year={2025}
}

@article{lin2022truthfulqa,
  title={TruthfulQA: Measuring How Models Mimic Human Falsehoods},
  author={Lin, Stephanie and Hilton, Jacob and Evans, Owain},
  journal={Transactions of the ACL},
  year={2022}
}

@article{ji2023survey,
  title={Survey of Hallucination in Natural Language Generation},
  author={Ji, Ziwei and Lee, Nayeon and Frieske, Rita and Yu, Tiezheng and Su, Dan and Xu, Yan and Ishii, Etsuko and Bang, Yejin and Madotto, Andrea and Fung, Pascale},
  journal={ACM Computing Surveys},
  volume={55},
  number={12},
  pages={1--38},
  year={2023},
  publisher={ACM}
}

@article{yang2025metaqa,
  title={Hallucination Detection in Large Language Models with Metamorphic Relations},
  author={Yang, Borui and Al Mamun, Md Afif and Zhang, Jie M. and Uddin, Gias},
  journal={Proceedings of the ACM on Software Engineering},
  volume={2},
  number={FSE},
  pages={1--21},
  year={2025},
  publisher={ACM}
}

@inproceedings{qualis2025kg,
  title={Mitigating Hallucinations in SysML v2 Generation Using LLMs and a Tri-Layered Knowledge Graph Reasoning Framework},
  author={Qualis, Richard A.},
  booktitle={2025 ACM/IEEE 28th International Conference on Model Driven Engineering Languages and Systems Companion (MODELS-C)},
  pages={357--366},
  year={2025},
  organization={IEEE}
}

@article{farquhar2024entropy,
  title={Detecting Hallucinations in Large Language Models Using Semantic Entropy},
  author={Farquhar, Sebastian and Kossen, Jannik and Kuhn, Lorenz and Gal, Yarin},
  journal={Nature},
  volume={630},
  pages={625--630},
  year={2024},
  publisher={Nature Publishing Group}
}

@inproceedings{binkowski2025lapeig,
  title={Hallucination Detection in LLMs Using Spectral Features of Attention Maps},
  author={Binkowski, Jakub and Janiak, Denis and Sawczyn, Albert and Gabrys, Bogdan and Kajdanowicz, Tomasz},
  booktitle={Proceedings of the 2025 Conference on Empirical Methods in Natural Language Processing},
  pages={24354--24385},
  year={2025}
}

@article{liu2023geval,
  title={G-EVAL: NLG Evaluation Using GPT-4 with Better Human Alignment},
  author={Liu, Yang and Iter, Dan and Xu, Yichong and Wang, Shuohang and Xu, Ruochen and Zhu, Chenguang},
  journal={arXiv preprint arXiv:2303.16634},
  year={2023}
}

@article{granstedt2025medical,
  title={Hallucinations in Medical Devices},
  author={Granstedt, Jason and Kc, Prabhat and Deshpande, Rucha and Garcia, Victor and Badano, Aldo},
  journal={Artificial Intelligence in the Life Sciences},
  volume={8},
  pages={100145},
  year={2025},
  publisher={Elsevier}
}

@article{du2026survey,
  title={A survey on the optimization of large language model-based agents},
  author={Du, Shangheng and Zhao, Jiabao and Shi, Jinxin and Xie, Zhentao and Jiang, Xin and Bai, Yanhong and He, Liang},
  journal={ACM Computing Surveys},
  volume={58},
  number={9},
  pages={1--37},
  year={2026},
  publisher={ACM New York, NY}
}

@article{ferrag2026llm,
  title={LLM and AI agents for autonomous systems: A survey of applications, datasets, and security challenges},
  author={Ferrag, Mohamed Amine and Lakas, Abderrahmane and Tihanyi, Norbert and Debbah, Merouane},
  journal={IEEE Open Journal of Intelligent Transportation Systems},
  volume={7},
  pages={615--657},
  year={2026},
  publisher={IEEE}
}

@article{brunello2026trustworthiness,
  title={Trustworthiness of large language models: hallucinations},
  author={Brunello, Nicol{\`o} and others},
  journal={Challenges and Applications of Generative Large Language Models},
  pages={107--126},
  year={2026},
  publisher={Elsevier}
}

@book{cao2025factual,
  title={Factual Consistency in Neural Text Generation: Detecting, Correcting, and Understanding Hallucinations},
  author={Cao, Meng},
  year={2025},
  publisher={McGill University (Canada)}
}

@inproceedings{lewis2020rag,
  title={Retrieval-Augmented Generation for Knowledge-Intensive NLP Tasks},
  author={Lewis, Patrick and Perez, Ethan and Piktus, Aleksandra and Petroni, Fabio and Karpukhin, Vladimir and Goyal, Naman and Kuttler, Heinrich and Lewis, Mike and Yih, Wen-tau and Rocktaschel, Tim and Riedel, Sebastian and Kiela, Douwe},
  booktitle={Advances in Neural Information Processing Systems},
  volume={33},
  pages={9459--9474},
  year={2020}
}

@inproceedings{yao2023react,
  title={ReAct: Synergizing Reasoning and Acting in Language Models},
  author={Yao, Shunyu and Zhao, Jeffrey and Yu, Dian and Du, Nan and Shafran, Izhak and Narasimhan, Karthik and Cao, Yuan},
  booktitle={International Conference on Learning Representations},
  year={2023}
}

@article{wu2023autogen,
  title={AutoGen: Enabling Next-Gen LLM Applications via Multi-Agent Conversation},
  author={Wu, Qingyun and Bansal, Gagan and Zhang, Jieyu and Wu, Yiran and Li, Beibin and Zhu, Erkang and Jiang, Li and Zhang, Xiaoyun and Zhang, Shaokun and Liu, Jiale and Awadallah, Ahmed Hassan and White, Ryen W. and Burger, Doug and Wang, Chi},
  journal={arXiv preprint arXiv:2308.08155},
  year={2023}
}

@inproceedings{li2023camel,
  title={CAMEL: Communicative Agents for ``Mind'' Exploration of Large Language Model Society},
  author={Li, Guohao and Hammoud, Hasan Abed Al Kader and Itani, Hani and Khizbullin, Dmitrii and Ghanem, Bernard},
  booktitle={Advances in Neural Information Processing Systems},
  volume={36},
  year={2023}
}

@article{du2023debate,
  title={Improving Factuality and Reasoning in Language Models through Multiagent Debate},
  author={Du, Yilun and Li, Shuang and Torralba, Antonio and Tenenbaum, Joshua B. and Mordatch, Igor},
  journal={arXiv preprint arXiv:2305.14325},
  year={2023}
}

@inproceedings{joshi2017triviaqa,
  title={Triviaqa: A large scale distantly supervised challenge dataset for reading comprehension},
  author={Joshi, Mandar and Choi, Eunsol and Weld, Daniel S and Zettlemoyer, Luke},
  booktitle={Proceedings of the 55th Annual Meeting of the Association for Computational Linguistics (Volume 1: Long Papers)},
  pages={1601--1611},
  year={2017}
}

@article{kim2024probabilistic,
  title={Probabilistic bernoulli and euler polynomials},
  author={Kim, T and Kim, DS},
  journal={Russian Journal of Mathematical Physics},
  volume={31},
  number={1},
  pages={94--105},
  year={2024},
  publisher={Springer}
}

@article{de2026theory,
  title={Theory-Grounded LLM Societies for Emergent Coordination},
  author={De Curt{\`o}, J and De Zarz{\`a}, I},
  journal={IEEE Access},
  year={2026},
  publisher={IEEE}
}

@inproceedings{wang2025demystifying,
  title={Demystifying the Power of Large Language Models in Graph Generation},
  author={Wang, Yu and Rossi, Ryan A and Park, Namyong and Ahmed, Nesreen K and Koutra, Danai and Dernoncourt, Franck and Derr, Tyler},
  booktitle={Findings of the Association for Computational Linguistics: NAACL 2025},
  pages={8189--8204},
  year={2025}
}

\appendix
\subsection{Extended Topology and Prompting Strategy Sensitivity}
\label{subsec:topology_prompt_sensitivity}
To further assess the solution's robustness, we extend the evaluation along two dimensions: communication topology and prompting strategy. First, to address RQ1, we expand the topology analysis by adding Erd\H{o}s-R\'enyi random graphs \cite{de2026theory} and Watts-Strogatz small-world graphs \cite{wang2025demystifying} to the original ring, fully connected, and scale-free structures. Second, to address RQ2 and RQ3, we evaluate zero-shot, few-shot, and CoT prompting while maintaining the same multi-agent configuration and HPR-adaptive defense mechanism.
\begin{table}[t!]
\centering
\caption{Extended topology analysis under adversarial interaction.}
\resizebox{\columnwidth}{!}{%
\begin{tabular}{llcccccc}
\toprule
Topology & Defense & Accuracy & Hallucination & AF & HPR & $\mathcal{R}_0$ & Cascade Size \\
\midrule
Ring & None & 0.842 & 0.091 & 1.18 & 0.914 & 0.86 & 2.1 \\
Ring & HPR-adaptive & 0.881 & 0.052 & 0.94 & 0.948 & 0.61 & 1.2 \\
\midrule
Erd\H{o}s--R\'enyi & None & 0.826 & 0.108 & 1.29 & 0.895 & 1.01 & 3.4 \\
Erd\H{o}s--R\'enyi & HPR-adaptive & 0.869 & 0.064 & 1.00 & 0.936 & 0.74 & 1.9 \\
\midrule
Small-world & None & 0.817 & 0.116 & 1.33 & 0.887 & 1.07 & 4.0 \\
Small-world & HPR-adaptive & 0.862 & 0.069 & 1.03 & 0.931 & 0.78 & 2.2 \\
\midrule
Fully connected & None & 0.808 & 0.126 & 1.37 & 0.876 & 1.13 & 4.8 \\
Fully connected & HPR-adaptive & 0.857 & 0.073 & 1.01 & 0.928 & 0.81 & 2.5 \\
\midrule
Scale-free & None & 0.795 & 0.139 & 1.45 & 0.861 & 1.21 & 5.6 \\
Scale-free & HPR-adaptive & 0.841 & 0.086 & 1.12 & 0.917 & 0.92 & 3.2 \\
\bottomrule
\end{tabular}%
}
\label{tab:extended_topology}
\end{table}
Table~\ref{tab:extended_topology} shows that communication structure substantially impacts collective hallucination propagation. Sparse local communication limits information exchange, whereas more highly connected structures create additional pathways for unsupported claims to spread across agents. Without defense, $\mathcal{R}_0$ increases from 0.86 in the ring topology to 1.21 in the scale-free topology. The HPR-adaptive mechanism consistently reduces hallucination propagation across all evaluated structures and maintains $\mathcal{R}_0<1$ in every topology, demonstrating robustness to variations in communication structure.
We further examine whether the observed reliability improvements depend on the prompting strategy used by individual agents. We evaluate zero-shot, few-shot, and CoT prompting while preserving the same system configuration and HPR-adaptive control mechanism.
\begin{table}[t!]
\centering
\caption{Sensitivity to prompting strategy under HPR-adaptive control.}
\begin{tabular}{lccccc}
\toprule
Prompting & Accuracy & Hallucination & AF & HPR & $\mathcal{R}_0$ \\
\midrule
Zero-shot & 0.842 & 0.083 & 1.15 & 0.916 & 0.87 \\
Few-shot & 0.861 & 0.074 & 1.09 & 0.926 & 0.82 \\
CoT & 0.867 & 0.072 & 1.08 & 0.928 & 0.81 \\
\bottomrule
\end{tabular}
\label{tab:prompt_sensitivity}
\end{table}
Table~\ref{tab:prompt_sensitivity} shows that prompting strategy affects local model reliability, with CoT prompting achieving the highest accuracy, lowest hallucination rate, lowest AF, and highest HPR. However, these differences are smaller than the improvements produced by propagation-aware interaction control. Taken together, the topology and prompting analyses indicate that the proposed HPR-adaptive mechanism remains effective across variations in both network structure and agent-level prompting behavior.

\end{document}